\documentclass[aps,pre,amsmath,twocolumn,superscriptaddress,showpacs]{revtex4-2}
\usepackage{calc}
\usepackage{amsmath}
\usepackage{amsfonts}
\usepackage{amssymb}
\usepackage{graphicx}
\usepackage{hyperref}
\usepackage{soul}
\usepackage{mathtools}
\usepackage{verbatim}
\usepackage{epstopdf}
\usepackage{subfigure}
\usepackage{latexsym}
\usepackage{dcolumn}
\usepackage{epsf}
\usepackage{float}
\usepackage[table]{xcolor}
\usepackage{multirow}
\usepackage[toc]{appendix}

\usepackage{color} 

\usepackage{graphicx}
\usepackage{graphicx,epstopdf}
\usepackage{xcolor}
\usepackage{dcolumn}
\usepackage{bm}
\usepackage{mathrsfs} 
\usepackage{mathrsfs}

\usepackage{colortbl}
\usepackage[table]{xcolor}
\usepackage{amsthm}

\theoremstyle{definition}

\theoremstyle{remark}

\begin{document}
	
	\preprint{APS/123-QED}
	
	\title[Metastability by non-reciprocal adaptive couplings]{Metastability induced by non-reciprocal adaptive couplings in  Kuramoto models} 
	
\author{Sayantan Nag Chowdhury}
\affiliation{School of Science, Constructor University, P.O.Box 750561, 28725 Bremen, Germany}

\author{Hildegard Meyer-Ortmanns}
\email{hmeyerortm@constructor.university}
\affiliation{School of Science, Constructor University, P.O.Box 750561, 28725 Bremen, Germany}
\affiliation{Complexity Science Hub, Vienna, Austria}

	%
	\date{\today}
	
	\begin{abstract}
		Non-reciprocal couplings are frequently found in systems out-of-equilibrium such as neuronal networks. We consider generalized Kuramoto models with non-reciprocal adaptive couplings. The non-reciprocity refers to the type of couplings according to Hebbian or anti-Hebbian rules and to different time scales on which the couplings evolve. The main effect of this specific combination of deterministic dynamics is an induced metastability of anti-phase synchronized clusters of oscillators. Metastable switching is typical for neuronal networks and a characteristic of  brain dynamics. We analyze the metastability as a function of the system parameters, in particular of the size  and the network connectivity. The mechanism behind  sudden changes in the order parameters is individual oscillators which change their cluster affiliation from time to time, providing ``weak ties" between clusters of synchronized oscillators, where an individual oscillator may represent an entire brain area. The time series exhibit random features but arise from deterministic dynamics.
	\end{abstract}
	
	\maketitle

\section{Introduction}

To bridge the gap between behavioral and neural views of learning, neural analogues of behavioral modification paradigms have been postulated. An example of such an attempt was D. Hebb's suggestion in 1949 \citep{hebb1949organization} that when a cell $A$ repeatedly and persistently takes part in firing another cell $B$, $A's$ efficiency in firing $B$ is increased. This is a rule for synaptic plasticity. It amounts to an analogue of associative conditioning. It is then typical for neural network models to employ a presynaptic and a postsynaptic correlation rule for altering connectivity as a mathematical representation of Hebb's postulate. Inspired by Hebb's seminal work, Rosenblatt  proposed the perceptron \citep{rosenblatt1958perceptron} to establish a relation between biophysics and psychology. His goal was to predict learning curves from neurological variables and vice versa, neurological variables from learning curves. He proposed a quantitative statistical approach to the organization of cognitive systems. This was in 1958. Already in the seventies and eighties, many adaptive neural network theories were considered in which neuron-like adaptive elements behave as single unit analogues of associative conditioning. In particular, such an adaptive element was proposed in \citep{sutton1981toward} that respects facts of animal learning theory. It implements an essential feature of classical conditioning that the element learns to increase its response rate in anticipation of increased stimulation. 
Later many theoretical studies of learning in neural systems were based on the Hopfield model \citep{hopfield1982neural} and McCulloch-Pitts  units \citep{hayman1999mcculloch} or ``integrate-and-fire" models of neurons, coupled via synapses that accumulate the presynaptic activity into an increasing membrane potential of neurons which fire when a certain threshold is crossed \citep{tuckwell1988introduction}. The synaptic strength there is not fixed. \\
In a complementary approach, neuronal activity is modeled by periodic oscillators which replace the neurons or entire sub-populations of neurons such as cortical areas. The reason is that rhythmic activity is observed, for example, in central pattern generators, visual or olfactory systems \citep{gray1994synchronous}. When neural networks are described as networks of coupled oscillators, the role of relative spike timing is played by the individual oscillator phases. It was Winfree \cite{winfree1967biological} who advanced a  reduction of the oscillator dynamics to a pure phase dynamics if the coupling is assumed to be weak, he discovered that a transition to synchronization occurs when the coupling strength exceeds a certain threshold. Kuramoto then proposed the sine interaction term which made the model analytically solvable. This led to the proposal of Kuramoto's model of phase oscillators \citep{kuramoto2005self, kuramoto1984chemical}, for a later review see \citep{acebron2005kuramoto}. In its original version, globally coupled Kuramoto oscillators with nonidentical frequencies exhibit a second-order phase transition to a synchronized state as the coupling strength increases above a critical value. Various generalizations of the Kuramoto model have been considered afterwards.\\
One of the first studies of plasticity and learning in a network of coupled phase oscillators is the work of \citep{sompolinsky1986temporal} and somewhat later the work of \citep{seliger2002young}. Plasticity refers to the change in the synaptic weights in the context of neural networks, it is a special form of general  adaptation dynamics  for which the couplings adapt to the dynamics of the dynamical units assigned to the nodes of the network. Also beyond applications to neural networks, coupling adaptation in oscillatory systems leads to many interesting effects, since the level of description is abstract enough to allow different interpretations.  Insights from neural networks with adaptive couplings can be exploited to explain dynamical properties of multi-frequency clusters in power grid networks, described by Kuramoto-Sakaguchi phase oscillators with inertia \citep{berner2021adaptive}. The fundamental relation between power grid and neural networks is established in \citep{berner2021modelling}; according to this relation phase oscillator models with inertia correspond to a particular class of adaptive networks.
Modular structure on meso- and macroscales can emerge when adaptively reinforced synchronization competes with a constraint on nodes to establish connections with other units of the network \citep{gutierrez2011emerging}.  Adaptation is also considered for adaptive multiplex networks, see, for example, the work of \citep{zhang2015explosive, berner2020birth, sawicki2022modeling}. Adaptive coupling can have the interesting effect of inducing collective excitability and self-sustained bursting oscillations in globally coupled populations of non-excitable units \citep{ciszak2020emergent}.  For oscillatory neuronal populations with spike-timing dependent plasticity it is shown in \citep{popovych2013self} that for an optimal noise level the amount of synaptic coupling gets maximal, leading to noise-induced self-organized synaptic connectivity. Here, the application of noise counteracts desynchronization that would naively be expected  as result of noise.\\

\begin{figure*}[htbp]
	\centering
	\includegraphics[width=0.8\textwidth]{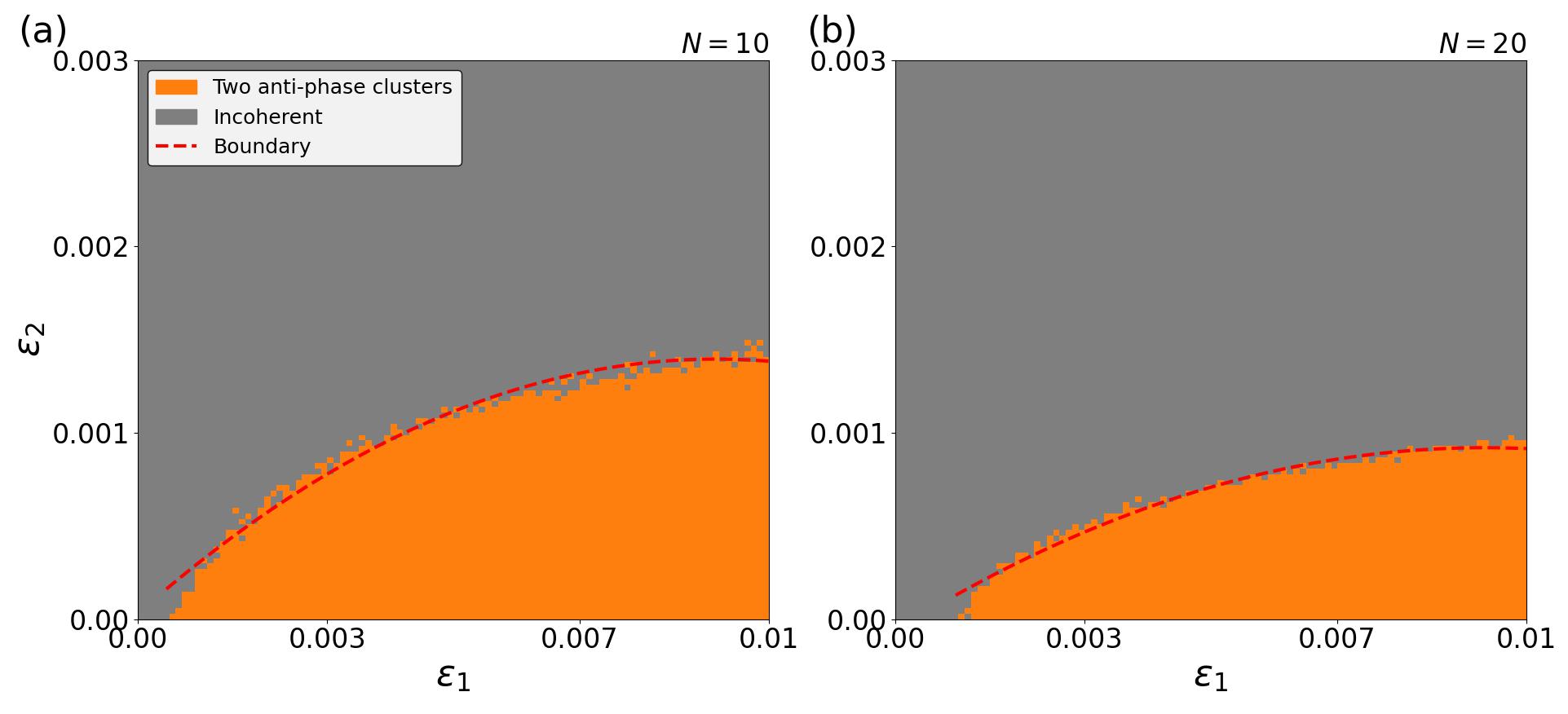}
	\caption{Categorical heatmap of collective states as a function of $(\varepsilon_1, \varepsilon_2)$ for (a) $N=10$ oscillators and (b) $N=20$ oscillators, averaged over $200$ independent realizations. For a specific choice of $\varepsilon_1,\varepsilon_2$ colors indicate  the state to which the system converges: orange for two anti-phase clusters, and gray for incoherent states; $\varepsilon_1$ is varying within [0, 0.01] and $\varepsilon_2$ is varying within [0, 0.003] with a step length of 0.0001. The red dashed curve serves to guide the eyes and roughly marks the transition boundary between two anti-phase clusters and incoherent states via a quadratic least square fit.
	}
	\label{fig1}
\end{figure*}

According to the work of \citep{kasatkin2017self}, a multilayer structure and chimera states emerge in a self-organized way if generalized globally coupled Kuramoto models are considered with adaptive couplings. Subnetworks of densely coupled elements form sequentially whose size is hierarchically ordered and which decouple as result of the hierarchical structure. If the global coupling in such networks is replaced by a random sparse structure of connections, chimera states form which retain features of a hierarchical organization, and the set of elements that form coherent groups can be rearranged during the network evolution \citep{kasatkin2018effect}. When the couplings adapt and have their own dynamics, it amounts to plasticity in relation to neuronal networks. In particular, the couplings may change as a function of the timing between the oscillators. If spike-time dependent plasticity also enters into the natural frequencies of the oscillators, heterogeneous layered clusters with different frequencies show up from homogeneous oscillator populations \citep{aoki2015self}. In this paper we will consider special cases of the work of \citep{kasatkin2017self, kasatkin2018effect, aoki2015self}. \\
More recently, possible biological or neurological mechanisms related to learning, plasticity,  and adaptation are reviewed  in \citep{papo2025biological}. Hebbian learning is one important mechanism of brain plasticity, but Hebbian learning alone would lead to dynamic instability and requires compensatory processes on multiple timescales \citep{zenke2017hebbian}. Various mechanisms are listed in \citep{papo2025biological} to achieve dynamic stability. It seems still to be relatively unknown at which spatial and temporal scales Hebbian, homeostatic, and other plasticity mechanisms interact, and which functional role they play. In this work, we will consider two plasticity mechanisms, Hebbian and anti-Hebbian, acting at quite different time scales, but this choice is not motivated by neurological experiments, it serves to explore the effect of non-reciprocity. \\ \\
At another and very different frontier, non-reciprocity is topical in systems out-of-equilibrium, since the validity of Newton`s third law (for every action is an equal and opposite reaction) is the exception rather than the rule in these systems. Arguably the most plausible example is from social systems where a relation of friendship holds in one direction but more enmity than friendship in the opposite direction. Populations with  conformist and contrarian members are considered in \citep{hong2011kuramoto}. Non-reciprocity is also found in  active matter \citep{lavergne2019group, uchida2010synchronization, nagy2010hierarchical}, metamaterials \citep{brandenbourger2019non, miri2019exceptional}, or in relation to game theory \citep{nag2020cooperation}. Occasionally it is  discussed in neuronal networks \citep{sompolinsky1986temporal, montbrio2018kuramoto}. In \citep{sompolinsky1986temporal}, asymmetric networks are considered, where only part of the synaptic connections between pairs of neurons are non-reciprocal and have slow dynamic response with the effect that temporal association becomes possible as recall of time sequences and cycles of patterns. The work of \citep{montbrio2018kuramoto} introduces two populations of excitatory and inhibitory populations in a Kuramoto model with the non-reciprocal feature that each oscillator of the excitatory (inhibitory) population exerts a positive (negative) influence on each oscillator of the inhibitory (excitatory) population, respectively. This way, the Kuramoto model accounts for the onset of excitatory/inhibitory-based neuronal rhythms.\\
Particularly in view of collective behavior, non-reciprocity can have strong implications on many-body systems. It may lead to time-dependent phases in which spontaneously broken continuous symmetries are dynamically restored \citep{fruchart2021non}. In view of general manifestations of non-reciprocity, we consider systems with self-organizing structures, of which synchronization is a prototypical form. More specifically, this paper is on non-reciprocity in adaptive networks and its impact on synchronization patterns. The dynamics of non-reciprocal adaptation may be realized in many ways, via the type of favored interaction (alignment or anti-alignment, Hebbian or anti-Hebbian rules), via the choice of time scales, via feedback as in \citep{sompolinsky1986temporal} or frequency adaptation to name some of them.\\
In this paper, we focus on generalized Kuramoto models with both adaptive and non-reciprocal couplings. When we combine Hebbian and anti-Hebbian adaptation rules for the couplings, acting on quite different time scales, we observe switching dynamics between metastable states, characterized  by anti-phase cluster synchronization (exceptionally also full synchronization) unless the choice of parameters leads to incoherent oscillations. Emerging metastability  is the main novel feature of our results. Therefore, the results may be relevant to neural network applications, since
many experimental hints exist that typical states in the brain are metastable \citep{tognoli2014metastable}. The structure of these metastable states is reflected in functional neuroimaging experiments. In particular cognitive phenomena rely on transient dynamics such as working memory and decision making \citep{rabinovich2008transient, rabinovich2011robust}. The idea is that brain activity is organized in spatiotemporal patterns through transient metastable states \citep{kelso1995dynamic, friston1997transients}. For a long time the search was for a dynamic framework that describes transient reproducible and metastable dynamics. One such framework is provided by heteroclinic dynamics, for a recent review see \citep{meyer2023heteroclinic}. For a review on metastable dynamics see, for example, \citep{rossi2025dynamical}.\\
In this paper, we demonstrate  how -differently from  heteroclinic dynamics- a set of deterministic equations can also lead to switching dynamics, here induced by a fast change  between attractive and repulsive couplings in half of the node connections, and a slow change between both types in the other half. In detail we consider an all-to-all coupling topology, but we also present results for a random network, for which we vary the density of connectivity. We briefly consider effects of non-reciprocal adaptivity when also the natural frequencies become coupling dependent.\\
The paper is organized as follows: In section 2 we introduce different versions of the model. Section 3 presents the results in terms of synchronization patterns as function of the model parameters, in particular of the time scales, the system size, and the grid topology. Section 4 contains our conclusions. In the appendix we provide an analytical understanding of the increasing instability with the system size.

\begin{figure*}[htbp]
	\centering
	\includegraphics[width=1.0\textwidth]{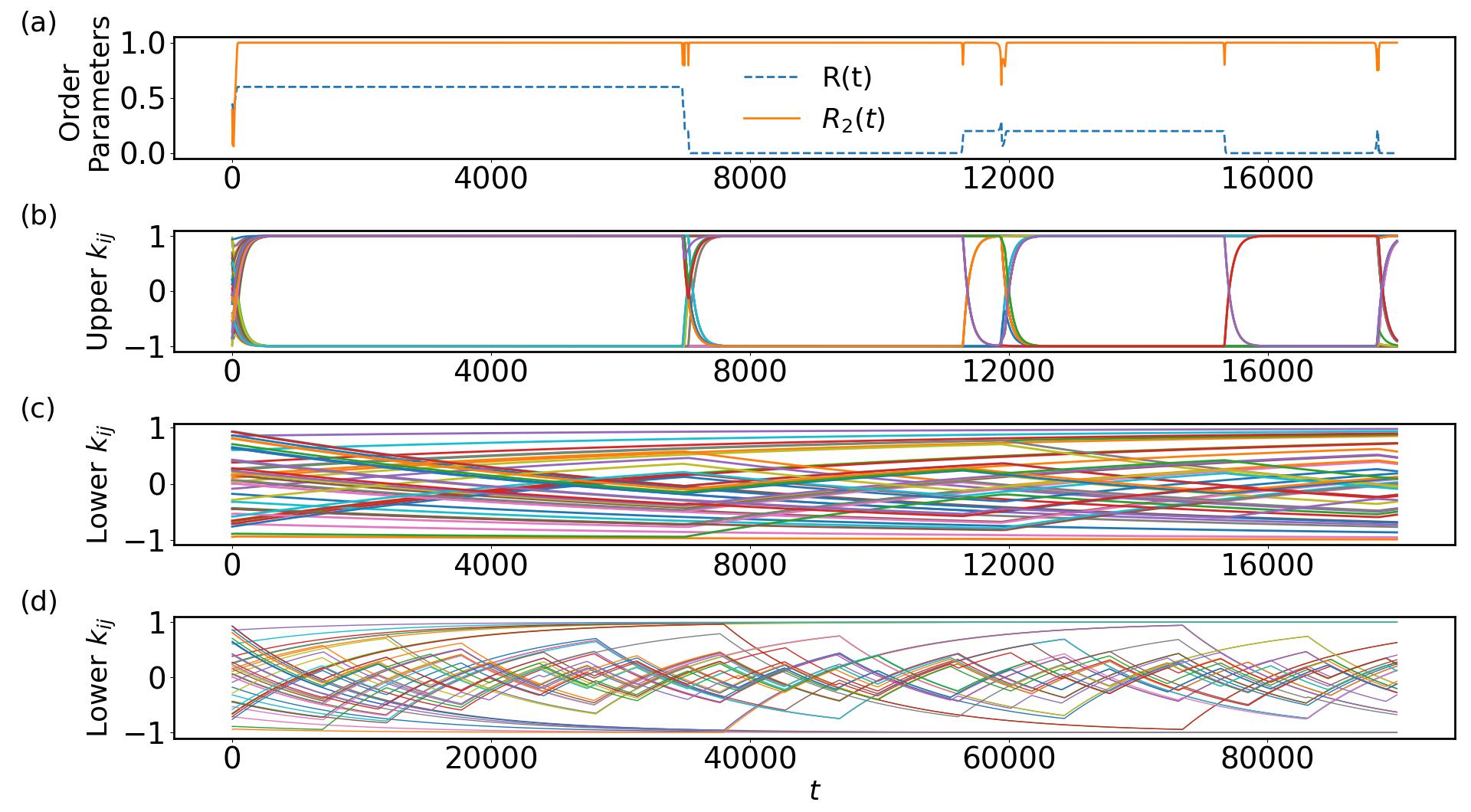}
	\caption{ (a) Time series of the first- and second-order Kuramoto order parameters $R(t)$ (dashed line) and $R_2(t)$ (solid line) for $\varepsilon_1=0.01$ and $\varepsilon_2=0.0001$ with $N = 10$. After initial convergence to $R_2 \approx 1$, occasional deviations occur, visible in spikes, which subsequently return to the same state. (b-c) Upper part of the couplings \( k_{ij} \) (for \( i < j \)) and lower part of \( k_{ij} \) (for \( i > j \)) . As soon as \( R_2 \) deviates from 1, a few oscillators switch clusters, which leads to changes in the slopes of the corresponding \( k_{ij} \) values. Since \( \varepsilon_1 = 0.01 \) is significantly larger than \( \varepsilon_2 = 0.0001 \), the upper \( k_{ij} \) values quickly reach their equilibrium values of \( \pm 1 \), whereas the lower \( k_{ij} \) values adjust more slowly. (d) Extended time view of (c) to see some structure in the evolution of the lower couplings.}
	\label{fig2}
\end{figure*}

\section{The Model}
We consider a population of $N \geq 2$ phase oscillators, each described by its phase $\theta_i(t) \in [0, 2\pi)$, whose dynamics evolve according to a modified Kuramoto model with adaptive non-reciprocal couplings. The evolution equation for the $i$-th oscillator is given by
\begin{equation}
	\dot{\theta}_i = \omega_i + \frac{1}{N} \sum_{j=1}^{N} k_{ij} \sin(\theta_j - \theta_i) + \xi_i(t),
	\label{eq:theta_dynamics}
\end{equation}
where $\omega_i$ denotes the intrinsic natural frequency, and $k_{ij}$ is the time-dependent coupling strength, with $i,j =1,...,N$. The coupling is all-to-all and directed, and the matrix elements $\mathbf{k} = [k_{ij}]$ evolve as
\begin{equation}
	\small
	\dot{k}_{ij} =
	\begin{cases}
		-\varepsilon_1 \left[k_{ij} + \sin(\theta_i - \theta_j + \beta_1)\right], & \text{for } i < j, \\
		-\varepsilon_2 \left[k_{ij} + \sin(\theta_i - \theta_j + \beta_2)\right], & \text{for } i > j, \\
		0, & \text{for } i = j.
	\end{cases}
	\label{eq:k_dynamics}
\end{equation}
We choose $\beta_1 = - \frac{\pi}{2}$ and $\beta_2 = + \frac{\pi}{2}$. The parameters $\varepsilon_1, \varepsilon_2 > 0$ set the different time scales for the adaptation rates for the upper and lower triangular parts of the matrix. The diagonal elements of the coupling matrix, $k_{ii}$, are set to zero to avoid self-coupling.
This means that for connections from a lower-index oscillator $i$ to a higher-index oscillator $j$, the coupling tends to adjust with a phase lag of $-\pi/2$ (Hebbian plasticity rule, where the weights for a pair of in-phase (or anti-phase) oscillators will increase (or decrease), respectively,) see, for example, \cite{aoki2015self}. For connections in the opposite direction, the adjustment follows a phase lead of $+\pi/2$ (anti-Hebbian rule), discussed, for example, in \cite{kasatkin2017self}. For simplicity the upper-triangular (lower-triangular) couplings $k_{ij}$ with $i<j$ ($i>j$) are termed the upper (lower) couplings in the following. 
The term $\xi_i(t)$ represents additive Gaussian white noise of intensity $\sigma$:
\begin{equation}
	\langle \xi_i(t) \rangle = 0, 
	\langle \xi_i(t) \xi_j(t') \rangle =  \sigma^2 \delta_{ij} \delta(t - t').
\end{equation}
Throughout the paper we will keep the slow (fast) time scale associated with the anti-Hebbian (Hebbian) rule, respectively, since otherwise we observe only incoherent oscillations. Our motivation for this choice does not directly come from experiments, it is to test the effect of non-reciprocity when it is implemented in the adaptation dynamics by the type and the time scales of adaptation. Moreover, for $\epsilon_1=\epsilon_2$ it allows a comparison  to the results of the work of \citep{berner2019hierarchical, kasatkin2017self, kasatkin2018effect, aoki2015self} for special choices of their parameters. For example, when the phase dynamics includes an additional parameter $\alpha$ to simulate a phase lag of the interaction, we set it to zero unless otherwise stated.
In previous studies of \citep{berner2023adaptive}-\citep{wang2020chimeras}, 
the adaptation rules were chosen symmetric with $\varepsilon_1=\varepsilon_2>0$. In these cases, if the phase shift was $\beta_1 = \beta_2 = +\pi/2$, the system typically remained incoherent \citep{kasatkin2017self}, while if the phase shift was $\beta_1 = \beta_2 = -\pi/2$ , two anti-phase clusters could form \citep{kasatkin2017self}. \\
As a stochastic extension of the standard Euler method, Eqs.~\eqref{eq:theta_dynamics} and \eqref{eq:k_dynamics} are integrated numerically using the Euler--Maruyama scheme with a fixed time step $\Delta t = 0.01$ and $\theta_i \mapsto \theta_i \bmod 2\pi$.\\
The initial phases $\theta_i(0)$ are sampled uniformly from $[0, 2\pi)$, while (unless  stated otherwise) the initial couplings $k_{ij}(0)$ are drawn from a uniform distribution in $[-1,1]$ with $k_{ii}(0) = 0$. The natural frequencies $\omega_i$ are in most cases set to zero, for some test runs (not presented) we sampled them from a normal distribution with a fixed mean $\omega_{\mathrm{mean}}$ and a fixed standard deviation $\omega_{\mathrm{std}}$.\\
The degree of phase coherence is quantified using two complex quantities, the first Kuramoto order parameter $R$ \citep{kuramoto1984chemical, acebron2005kuramoto, daido1992order} and the second Kuramoto order parameter $R_2$ \citep{daido1992order}:
\begin{equation}
	R e^{i \psi} = \frac{1}{N} \sum_{j=1}^{N} e^{i \theta_j}, \qquad R_2 e^{i \psi_2} = \frac{1}{N} \sum_{j=1}^{N} e^{2i \theta_j}, \label{eq:R2}
\end{equation}
where $R \in [0,1]$ measures the global synchronization level, and $R_2 \in [0,1]$ captures the presence of two anti-phase clusters \citep{berner2019hierarchical}. It is particularly useful to characterize  our synchronization patterns. The limiting cases are as follows: $R \approx 1$ and $R_2 \approx 1$ for complete in-phase synchronization; $R \approx 0$ and $R_2 \approx 1$ for two anti-phase clusters; and both $R$ and $R_2$  small for incoherent dynamics. \\

\begin{figure*}[]
	\centering
	\includegraphics[width=\linewidth]{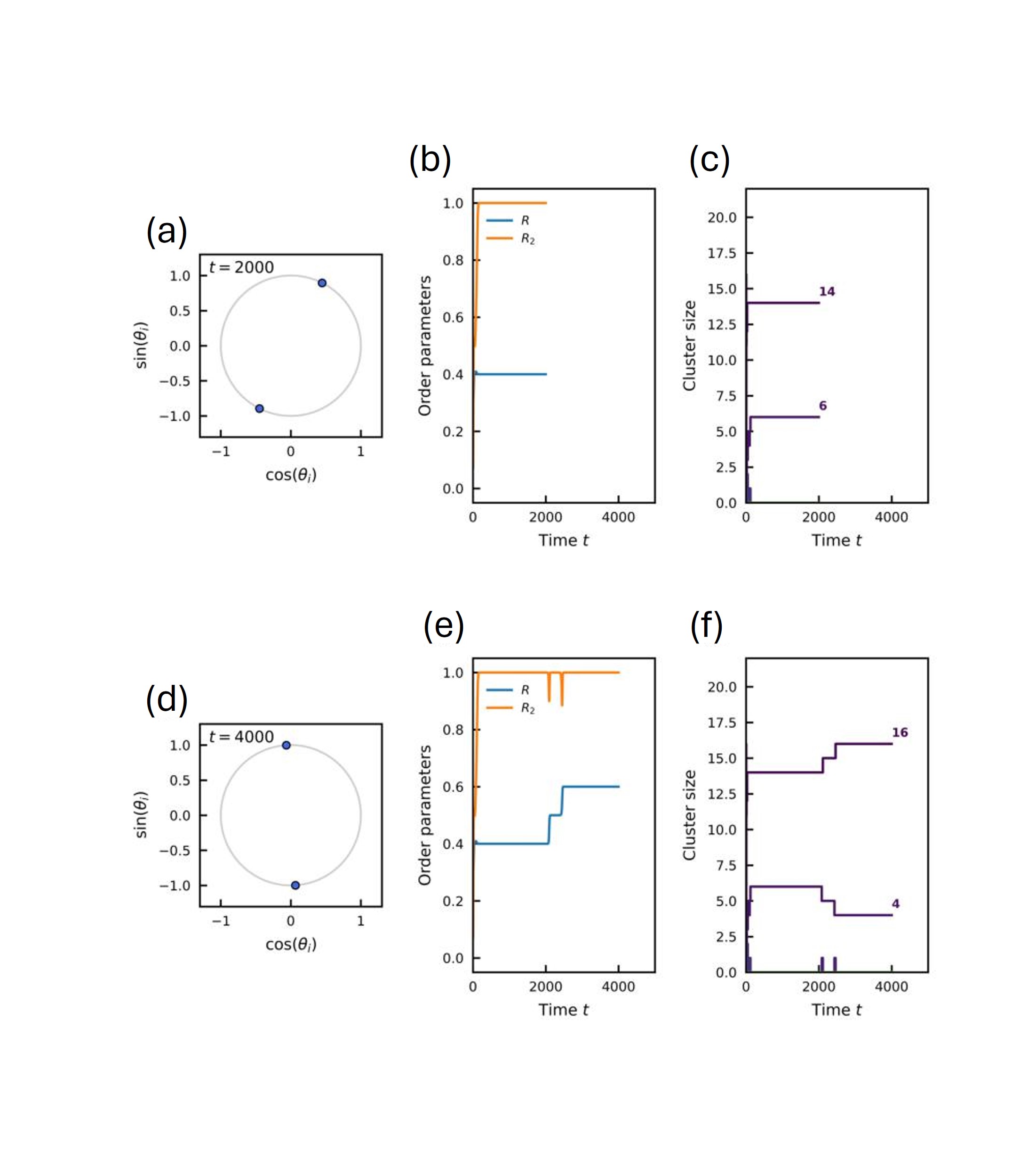}
	\caption{Snapshots of the system state before and after a switching event for $N=20$. \textbf{(a--c)} State at $t=2000$ (before the switch). (a) Instantaneous phases on the unit circle showing two well-separated clusters. (b) Time evolution of the order parameters $R$ and $R_2$. (c) Time evolution of the cluster sizes, showing a partition into $14$ and $6$ oscillators. \textbf{(d--f)} State at $t=4000$ (after the switch). (d) Instantaneous phases. (e) Evolution of order parameters; note the jump in $R$ and $R_2$ coinciding with the switch. (f) Cluster sizes reorganize into a partition of $16$ and $4$ oscillators.}
	\label{fig:switch_N20}
\end{figure*}

After an initial transient, the long-time averaged order parameters $\langle R \rangle$ and $\langle R_2 \rangle$ are computed over a sufficiently large portion of the simulation succeeding the transient. Unless otherwise specified, we simulate the system for a total of $T_{total} = 1.8 \times 10^6$ time steps. To eliminate initial transient behaviors, we discard the first $30\%$ of the simulation (approx. $5.4 \times 10^5$ steps) before computing the long-time averaged order parameters $\langle R \rangle$ and $\langle R_2 \rangle$. As we will see, for our results it is sufficient to classify the collective states as follows: if  \( \langle R \rangle \) and  \( \langle R_2 \rangle \) both are greater than a threshold \( R_{\mathrm{th}} \), the system is in the in-phase synchronized state. If \( \langle R \rangle \) is less than or equal to \( R_{\mathrm{th}} \), and \( \langle R_2 \rangle \) is greater than \( R_{\mathrm{th}} \), the system is in the two anti-phase clusters state. In all other cases, the system is considered incoherent. For the threshold we have chosen $R_{\mathrm{th}} = 0.8$, though usually for anti-phase synchronized clusters $R_2>0.8$. The threshold of $R_{th}=0.8$ acts as a conservative lower bound \citep{rodrigues2016kuramoto,arenas2008synchronization}. In the metastable states that we observe, the synchronized clusters are typically tightly locked, yielding $R_2$ values well above $0.9$. A threshold of $0.8$ reliably separates these coherent states from incoherent regimes while allowing for small fluctuations induced by the adaptive, non-reciprocal coupling dynamics.

\begin{figure*}[htbp]
	\centering
	\includegraphics[width=0.8\textwidth]{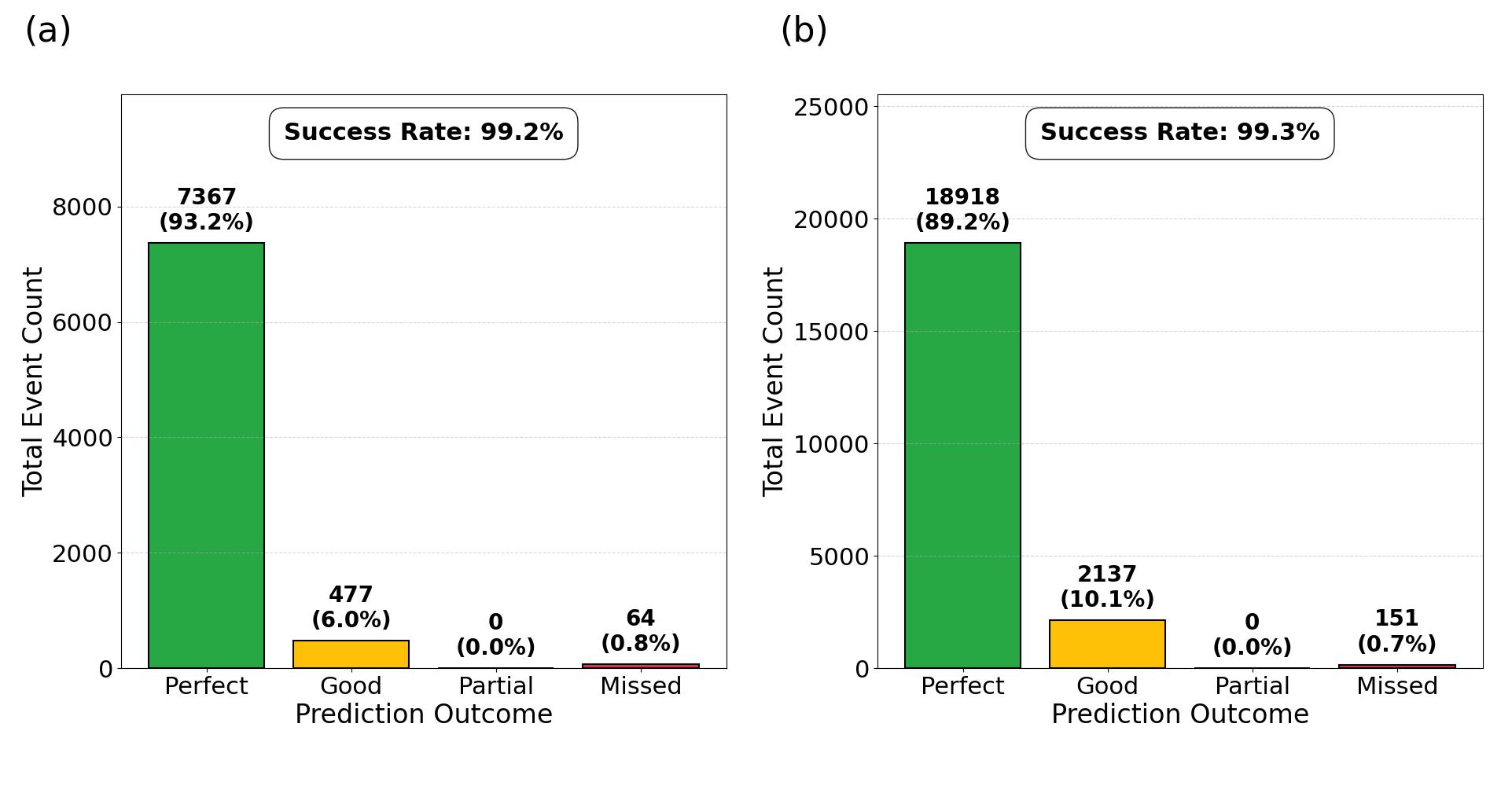}
	\caption{Snapshot-based prediction performance across $1,000$ random realizations for (a) $N=6$ and (b) $N=10$. The bars represent the total count of switching events categorized by prediction accuracy. The percentage shown above each bar indicates the frequency of that outcome relative to the total number of detected events. The overall success rate (combined \textit{Perfect} and \textit{Good} categories) remains remarkably stable as the system size increases.}
	\label{fig5}
\end{figure*}

\section{Results}
We start with a discussion of the typical stationary states and argue for the choice of parameters which we keep throughout this paper. Next we analyze features of randomness in spite of deterministic dynamics, the dependencies on the initial conditions, the system size, and the effect of disorder. In the last two subsections we consider different versions of the generalized Kuramoto model. 

\subsection{Stationary states and choice of parameters}
\label{sec:categorical_heatmap}

To visualize the dependence of collective states on the adaptation rates $(\varepsilon_1, \varepsilon_2)$, we perform  numerical simulations of the coupled system \eqref{eq:theta_dynamics}-\eqref{eq:k_dynamics} with $N=10$ and $N=20$ oscillators over multiple realizations.\\
The simulations are carried out in the noise-free case with identical intrinsic frequencies $\omega_i = 0$ for all oscillators $i = 1, \dots, N$. 
To track the long-time dynamics, the order parameters, $R$ and $R_2$, are computed and averaged over the last sufficiently long time window. Each parameter pair $(\varepsilon_1, \varepsilon_2)$ is simulated for $200$ independent realizations, and the resulting mean order parameters $\langle R \rangle$ and $\langle R_2 \rangle$ are used to classify the collective state.  Specifically for the construction of these phase diagrams, where covering the parameter space with high resolution requires massive computational throughput ($200$ realizations per grid point), we used a shorter simulation time of $4 \times 10^4$ steps. The order parameters were averaged over the final $1000$ steps after the system had relaxed into its attractor. We verified that this duration is sufficient for the collective state (two anti-phase clusters vs. incoherent) to stabilize.

The resulting categorical heatmap is shown in Fig.~\eqref{fig1}. Each pixel corresponds to an $(\varepsilon_1, \varepsilon_2)$ pair, color-coded by the observed collective state. Since $\varepsilon_2>0$ corresponds to the anti-Hebbian coupling, it tends to disrupt phase coherence, and sufficiently large values of $\varepsilon_2$ generally lead to incoherent dynamics. In contrast, when $\varepsilon_1 \gg \varepsilon_2 > 0$, the upper-triangular Hebbian interactions dominate, promoting the formation of two anti-phase clusters for most parameter realizations. For $\varepsilon$-values in the vicinity of the red dashed line it depends on the random initial conditions in which state the system ends up, therefore the boundary is a bit rough.\\
To examine the dynamics in detail, we simulate the system \eqref{eq:theta_dynamics}-\eqref{eq:k_dynamics} with $N=10$ oscillators, noise intensity \( \sigma = 0 \) and intrinsic frequencies $\omega_i = 0$ $\forall i$. We fix the coupling parameters at $\varepsilon_1 = 0.01$ and $\varepsilon_2 = 0.0001$, a regime where two anti-phase clusters are expected from Fig.~\eqref{fig1} (a). While one might anticipate that $R_2$ would converge to one after an initial transient, we observe metastable behavior: $R_2$ initially approaches $1$, then occasionally deviates from this value, only to return to $R_2 \approx 1$ after a few iterations. We observe the same recurrent deviations of $R_2$ as shown in Fig.~\eqref{fig2} (a) also over a total integration time of up to $9 \times 10^7$ steps.\\
As seen in Fig.~\eqref{fig2} (a), also the standard Kuramoto order parameter $R$ shows metastable behavior. Depending on the degree of synchronization, the value of $R$ after a spike may be different, and whenever $R_2$ is perturbed, $R$ experiences a corresponding disturbance. 
During $R_2=1$, the two anti-phase clusters remain well separated by a phase difference of $\pi$, although the number of oscillators in each cluster may vary temporarily during the disturbances of $R_2$. After a spike in $R_2$, the system returns to a state with $R_2 \approx 1$, preserving the two-cluster structure with $\pi$ phase difference. Thus, a spike in the order parameter $R_2$ indicates a short deviation from the organization into two anti-phase synchronized clusters which obviously corresponds to an unstable solution. 

\noindent
It is  instructive to plot  the evolution of the asymmetric coupling matrix \( k_{ij} \) in Fig.~\eqref{fig2} (b-c).  
In panel (b), we observe the time evolution of the upper  part  \( k_{ij} \), that is, for \( i < j \). As the system evolves, some of the couplings settle temporarily at \( +1 \), while others stabilize temporarily at \( -1 \). However, when \( R_2 \) deviates from \( 1 \), there are switches in some of the \( k_{ij} \) values from \( +1 \) to \( -1 \) or vice versa, indicating a temporary departure from the anti-phase synchronized two-cluster state.
At the same time, some of the lower couplings \( k_{ij} \), that is, for \( i > j \) also undergo slope changes, as shown in panel (c), but evolving on a much slower time scale. Note that the transient dynamics of the order parameters $R$ and $R_2$ appear slightly faster even than those of the upper couplings $k_{ij}$ (Fig.~\ref{fig2} (a-b)). This is due to the timescale separation explicitly built into the model: the phases $\theta_i$ evolve on a timescale of $O(1)$ (Eq.~ \eqref{eq:theta_dynamics}), whereas the upper part of the couplings evolve with rate $\varepsilon_1 = 0.01$. Consequently, the oscillators can phase-lock and stabilize $R$ and $R_2$ relatively quickly, while the coupling weights require some integration time to saturate to their equilibrium values. Panel (d) displays the evolution of the lower couplings on a long time scale, where it becomes more clearly visible that not all couplings are affected by a switch of $R_2$. Here one may wonder whether some periodic structure is seen on long time scales.

\noindent {\bf Metastability as an essential novel feature.} Metastability is the essential novel feature that we observe in these combinations of non-reciprocal couplings with respect to their sign and time scales. It is absent in the reciprocal case \citep{kasatkin2017self}. Even, when \( \varepsilon_1 \gg \varepsilon_2 \) with \( \varepsilon_1 = 0.01 \), \( \varepsilon_2 = 0.0001 \), but \( \beta_1 = \beta_2 \), the system does not exhibit metastability. In this case, when \( \beta_1 = \beta_2 = - \frac{\pi}{2} \), two stable  anti-phase clusters emerge, while for \( \beta_1 = \beta_2 = + \frac{\pi}{2} \), oscillations are only incoherent. Metastability, on the other hand, is observed in our system when there is sufficient asymmetry in the adaptive coupling dynamics, specifically when \( \varepsilon_1 \gg \varepsilon_2 > 0 \), as shown in the orange-colored region of Fig.~\eqref{fig1} as well as $\beta_1 \neq \beta_2$. The slow dynamics acts as a perturbation on the fast dynamics; for the fast dynamics alone we would see a kind of dynamic realization of Harary's theorem \citep{harary1960matrix}. According to that, the set of all points of a balanced signed graph can be partitioned into two disjoint sets such that each positive line joins two points of the same subset and each negative line joins two points from different subsets, the lines are undirected. In our case, we have directed links with different dynamics assigned to both directions; but as long as the lower couplings are neglected which act in opposite link direction to the upper ones, what dynamically matters is which dynamics is assigned to the lines: these are attractive upper couplings, corresponding to positive lines, and realized as zero phase difference  within the same cluster, as well as repulsive upper couplings, corresponding to negative lines, and realized as phase difference of $\pi$ between members of different clusters. The two anti-phase synchronized clusters are then slightly perturbed by the slow dynamics of lower couplings which is manifest in some oscillators changing their cluster affiliation from time to time.

\noindent {\bf Switching the cluster affiliation.} Numerically we confirm that occasionally some oscillators switch groups, causing \( R_2 \) to deviate from 1. After a few steps, these oscillators return to a possibly slightly different anti-phase cluster configuration, and \( R_2 \) revisits the value of 1. The number of oscillators in each cluster may change with each deviation of \( R_2 \) from 1. Here, we include four videos \citep{Sayantan2025Metastability} for different system sizes, \( N = 20 \) and \( N = 200 \), demonstrating these features, in particular how the group sizes may change with each spike in \( R_2 \). In Fig.~\eqref{fig:switch_N20}, we include snapshots of these movies before  and after a spike in $R_2$.

\begin{figure*}[htbp]
	\centering
	\includegraphics[width=0.8\textwidth]{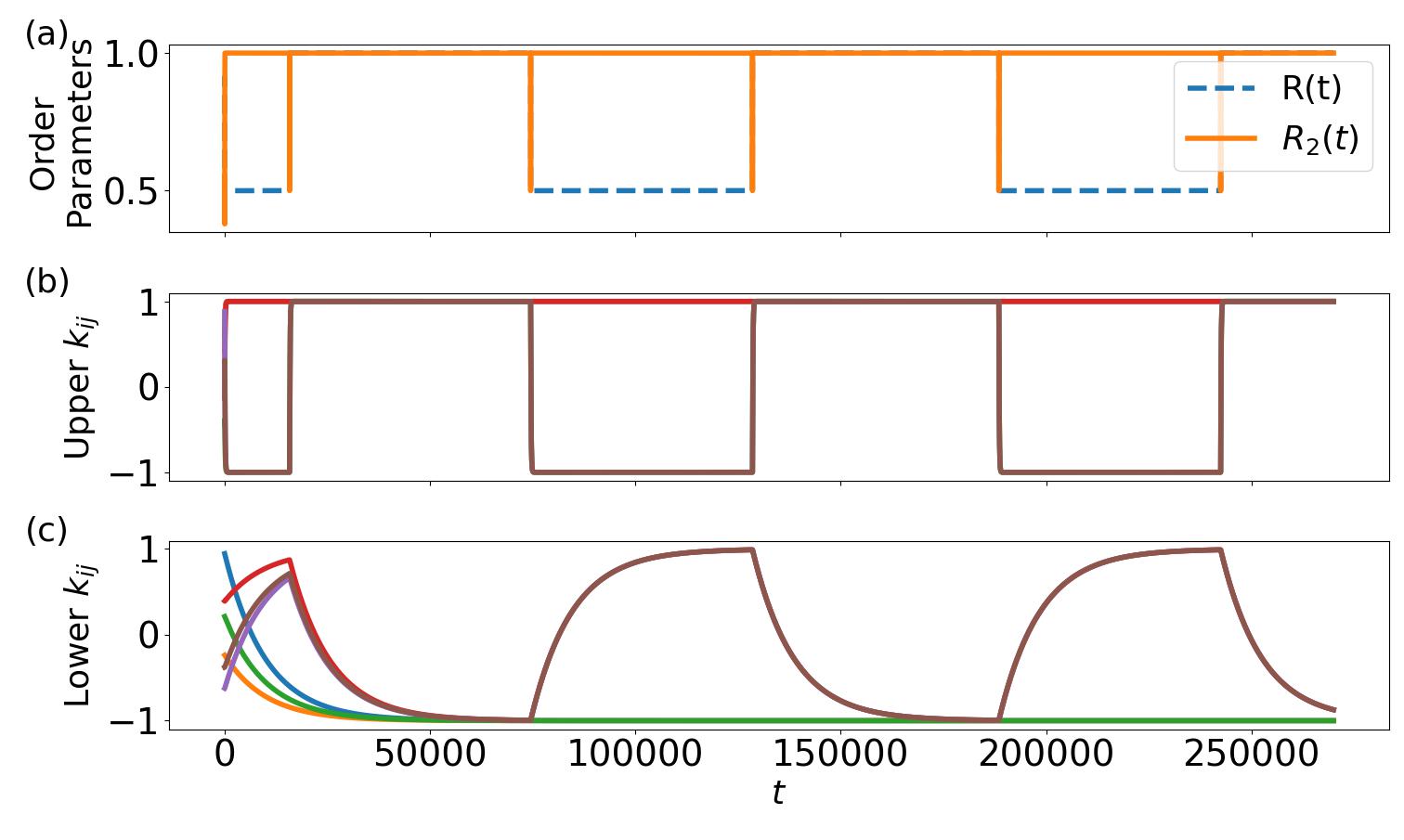} 
	\caption{Periodic and synchronized  evolution not only of phases (a), but also of upper and lower couplings into one or two clusters for a network of $N=4$ oscillators with only one saddle. Other parameters as in Fig.~\ref{fig2}.}
	\label{fig6}
\end{figure*}

\noindent {\bf What causes the changes in the slope of fast and slowly evolving couplings and which couplings are affected?} Let us consider the simplest case, where a single oscillator $l$ changes its cluster affiliation. What happens to the couplings $k_{lk}$ in which this oscillator is involved? From the evolution of the coupling strengths \( k_{ij} \) given in Eq.\ \eqref{eq:k_dynamics} it is seen that 
two oscillators with indices \( l \) and \( k \), chosen from the \( N \) oscillators, which initially belong to the same cluster at time \( t = t_1 \) in an anti-phase cluster configuration with \( R_2 = 1 \) 
evolve as

\begin{align*}
	\dot{k}_{lk} &= -\varepsilon_1 \left( k_{lk} - 1 \right), &\text{for } l < k, \\
	\dot{k}_{lk} &= -\varepsilon_2 \left( k_{lk} + 1 \right), &\text{for } l > k.
\end{align*}

Since \( k_{lk} \in [-1,1] \) and \( \varepsilon_1, \varepsilon_2 > 0 \), these relations imply
$\dot{k}_{lk} \ge 0, \; \text{for } l < k,$ and $\dot{k}_{lk} \le 0, \; \text{for } l > k$.
Thus, when oscillators \( l \) and \( k \) are in the same cluster, the upper–triangle elements of the coupling matrix (\( l < k \)) increase toward \( +1 \), while the lower–triangle elements (\( l > k \)) decrease toward \( -1 \).  Now suppose that at time \( t = t_2 > t_1 \), oscillator \( l \) switches its affiliation to the opposite cluster, so that now \( \theta_l - \theta_k = \pi \). In this case, the coupling dynamics becomes

\begin{align*}
	\dot{k}_{lk} &= -\varepsilon_1 \left( k_{lk} + 1 \right), &\text{for } l < k, \\
	\dot{k}_{lk} &= -\varepsilon_2 \left( k_{lk} - 1 \right), &\text{for } l > k.
\end{align*}

and, because \( k_{lk} \in [-1,1] \), we obtain $
\dot{k}_{lk} \le 0, \;\; \text{for } l < k$, and $\dot{k}_{lk} \ge 0, \;\; \text{for } l > k.$

{In conclusion}, this analysis shows that as long as two oscillators are in the same cluster, which is a metastable configuration, the couplings \( k_{lk} \) evolve toward the saddle-equilibrium $\dot{k}_{lk} = 0$  with either positive or negative slope. However, when one of the oscillators switches to the opposite cluster, the sign of the phase difference changes by \( \pi \), and the slope of evolution of \( k_{lk} \) reverses. Due to the difference in time scales this happens fast for upper couplings and very slowly for lower ones, so slowly that the lower couplings do not reach the equilibrium values of  $\dot{k}_{lk} = 0$ before the next switch happens.\\ 
Numerically and visible in the attached movies we observe that depending on the system size, for small sizes $N=10$ or $N=20$, it is a single or a few oscillators that change their cluster affiliation at the spikes of $R_2$, for larger sizes these are more.
What is typical in the movies is the splitting into two clusters of similar size as well as the fact that only a few oscillators change their affiliation. This reminds is reminiscent to ``weak ties" in social systems, where a few members of a community are weakly bound and willing to switch their community (of shared opinions, for example), while the majority is strongly bound, and the weak ties are of beneficial impact on the system, as analyzed by Granovetter \citep{granovetter1973strength}. The reason why oscillators change their cluster affiliation is determined by the attractor landscape and the saddles generated by the dynamics of Eqs.~(\ref{eq:theta_dynamics})-(\ref{eq:k_dynamics}) that we analyze in section~\eqref{subsechilde1}. In brain dynamics this would correspond to oscillatory units which easily switch the group with which they synchronize, where a single unit may represent  a population of neurons.

\begin{figure*}[]
	\centering
	\includegraphics[width=0.8\textwidth]{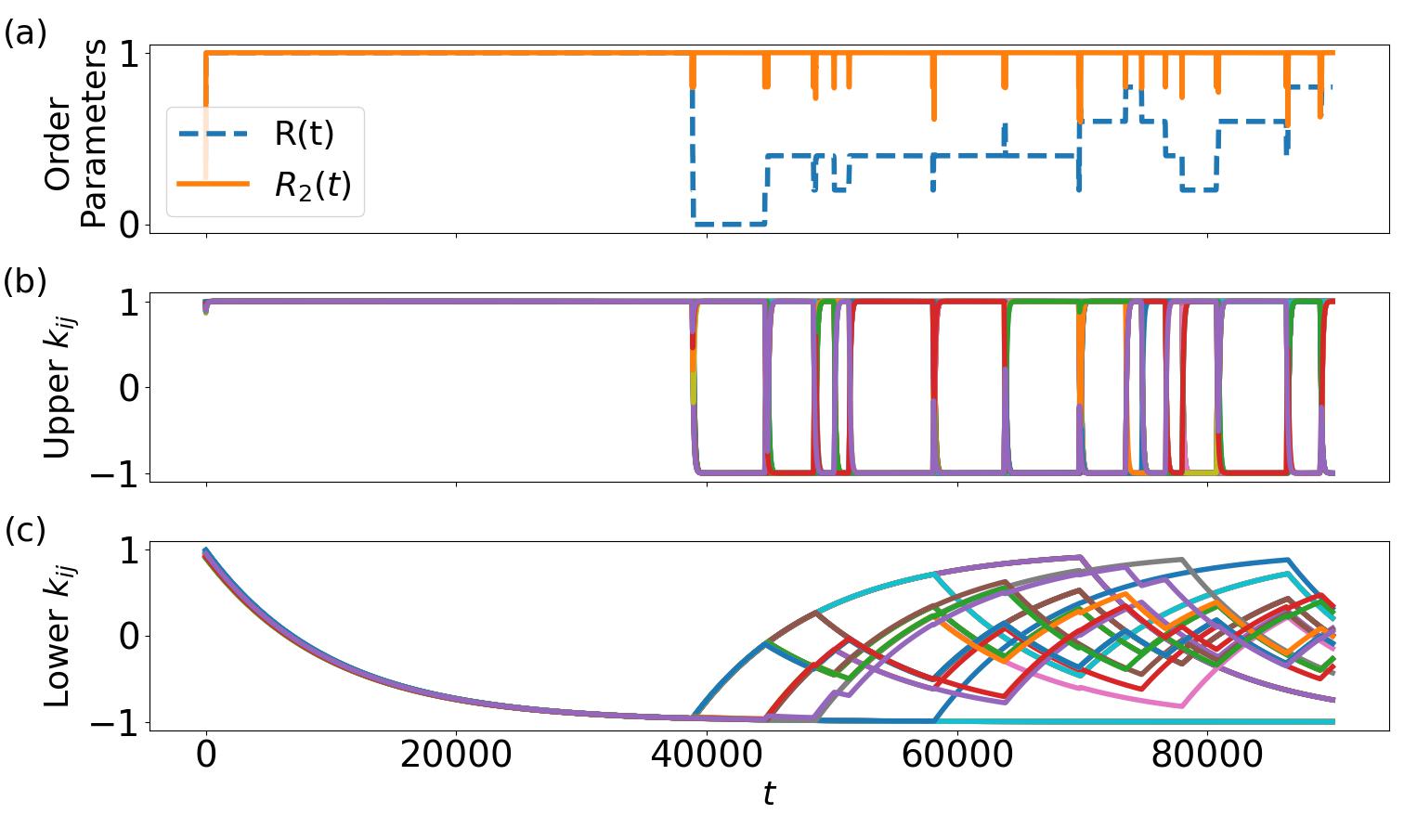}
	\caption{Dependence on the initial conditions for upper and lower couplings, here both chosen from a distribution of $k_{ij}$ close to $+1$. Note that the first long transient corresponds to a fully synchronized solution, observed very rarely as transient.
	}
	\label{fig7}
\end{figure*}

\subsection{Features of randomness}
The time evolution of the order parameters appears to be generated by a stochastic source in spite of the fully deterministic dynamics of Eqs.~\eqref{eq:theta_dynamics}-~\eqref{eq:k_dynamics}. The events when one or more oscillators change their cluster affiliation look randomly distributed. The only source of randomness is the quenched distribution of initial phases and couplings which determine the initial distance of the phases from the saddle equilibria and therefore their trajectories through attractor space. Actually, on an intermediate time scale of the order of the interval between two switching events, the distance from the next saddle is essential to predict which oscillator escapes first in a next switching event. Assuming that there is a switch at time $t$ of one or a few oscillators and that we know the  instantaneous phase velocities $\dot{\theta}_i$  at an earlier time $t_{snap} = t - \tau_{win}$  
from a snapshot at that  time, we approximate the estimated time of escape (ETE) for each oscillator $i$ as
\begin{equation}
	\eta_i = \frac{|\theta_{B} - \theta_i(t_{\text{snap}})|}{\dot{\theta}_i(t_{\text{snap}})\, dt}.
\end{equation}
Here $\theta_B$ is chosen as $\pi$. This choice is  motivated by the circular topology of the phase space: $\pi$ marks the point of maximum phase separation from the origin ($0$ or $2\pi$), and crossing it signals a definitive change in cluster membership, since we have two anti-phase synchronized clusters.  $\dot{\theta}_i(t_{\text{snap}})\, dt$ represents the phase displacement per simulation increment at the time of the snapshot, so the ratio on the right-hand side gives the time in units of simulation steps it takes oscillator $i$ to overcome distance $\theta_{B} - \theta_i(t_{\text{snap}})$. This time is minimal for the oscillator $i^*$ which is the first to change its cluster affiliation.
\begin{equation}
	i^* = \arg\min_i (\eta_i).
\end{equation}
Note that the snapshot of velocities $\dot{\theta}_i(t_{\text{snap}})$ depends on the upper and lower couplings which coevolve with the phases. However, our previous figures for $k_{ij}$ show that within two switching events upper couplings are almost constant, as they converge fast to their equilibrium values, and lower couplings are so slow that their change may be neglected. Moreover, the distance is assumed to increase linearly with time. Nevertheless this approximation works quite well to predict the dynamics after a switch including an identification which oscillator will switch first. \\
Before we discuss the concrete results, some remarks are in order about the choice of $\tau_{{\text{win}}}$, $\tau_{{\text{win}}}=200$ turns out to be a good choice, given the other parameters. If $\tau_{{\text{win}}}$ is chosen too large, the co-evolution of couplings can no longer be neglected, oscillators whose ETE exceeds a cutoff value $\eta_{\max} = 500$ steps are excluded from consideration. If $\tau_{{\text{win}}}$ is too short, there is no prediction. If more than one oscillator switches its cluster affiliation, we  group individual switches that happen very close together into a single collective event. If several oscillators overcome the distance of $\pi$ within $50$ simulation steps of each other, they are evaluated as a single event. For entire groups changing their affiliation, we predict the membership of the switching cluster $\mathcal{P}$ by including any oscillator $j$ whose ETE lies within a temporal margin $\Delta \eta$ of the ``leading oscillator" $i$:
$\mathcal{P} = \{ j \mid \eta_j \le \eta_{i^*} + \Delta \eta \}$.
This margin accounts for the finite temporal spread of collective switching events and is set to $\Delta \eta = 150$ steps in our simulations, based on the typical duration of observed group switching.\\
\noindent By comparing the predicted set $\mathcal{P}$ with the actual set of oscillators $\mathcal{A}$ that overcome a distance of $\pi$ at time $t$, predictions are classified into four  categories:
Perfect: $\mathcal{P} = \mathcal{A}$. The predicted group exactly matches the actual switchers.
Good: $\mathcal{A} \subset \mathcal{P}$. All actual switchers were identified, but additional oscillators were incorrectly included.
Partial: $\mathcal{P} \cap \mathcal{A} \neq \emptyset$ but $\mathcal{A} \not\subset \mathcal{P}$. Some switchers were correctly identified, while others were missed.
Missed: $\mathcal{P} \cap \mathcal{A} = \emptyset$. No actual switchers were identified.
To skip initial transient behavior, only switching events occurring after $t > 0.3\,T_{\text{total}}$ are included from overall $T_{\text{total}} = 1.8 \times 10^6$ steps.

\begin{figure*}[htbp]
	\centering
	\includegraphics[width=0.8\linewidth]{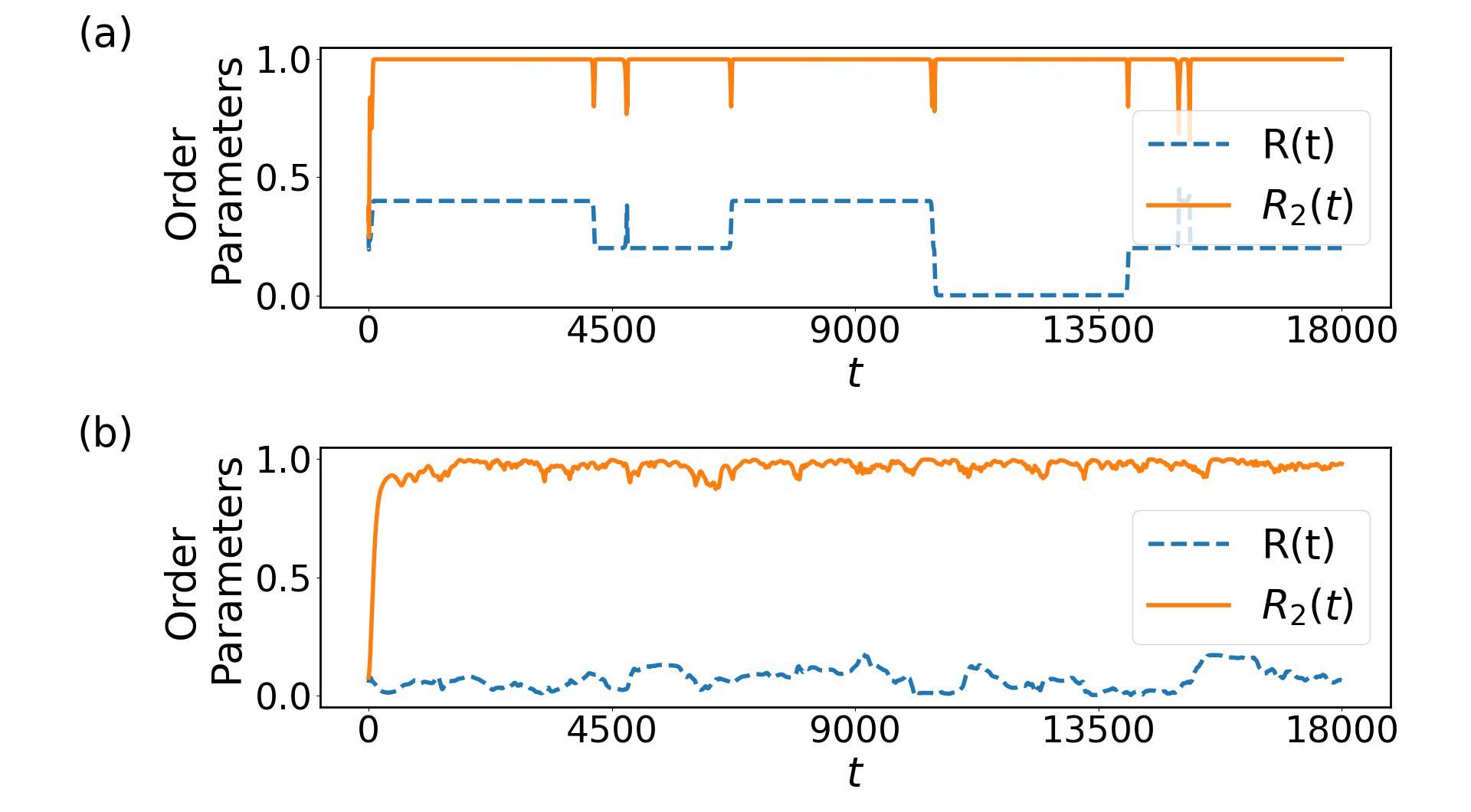}
	\caption{Dependence on the system size for (a) $N=10$ and (b) $N=200$ oscillators with the same parameters as in Fig.~\eqref{fig2}. Compared to the smaller system ($N=10$), the fluctuations in $R_2$ are more frequent, indicating a larger instability, but less pronounced in size, that is, smaller deviations from the anti-phase synchronized clusters for $N=200$.  Still, all values of $R_2$ remain close to one, corresponding to an organization into two anti-phase clusters.
	}
	\label{fig8}
\end{figure*}

\noindent The performance of our snapshot-based prediction is summarized in Fig.~\ref{fig5} for both $N=6$ and $N=10$ nodes. For the $N=6$ configuration (Fig.~\ref{fig5} (a)), the framework achieved an overall success rate of $99.2\%$, with $93.2\%$ of events classified as \textit{Perfect}. This indicates that the instantaneous state 200 steps prior to a switch contains nearly all the information required to identify the switching cluster. As the system size increases to $N=10$ (Fig.~\ref{fig5} (b)), the predictive robustness is maintained with a success rate of $99.3\%$. Interestingly, while the proportion of \textit{Perfect} predictions remains high at $89.2\%$, there is a slight increase in the \textit{Good} category ($10.1\%$ compared to $6.0\%$ in the $N=6$ case). This shift suggests that in larger coupled systems, the ``lead" oscillator exerts a stronger influence on its neighbors, occasionally causing the inclusion of `follower' oscillators in the predicted set $\mathcal{P}$ that do not quite reach $\theta_B$ within the expected window. The zero occurrence of \textit{Partial} predictions across all realizations demonstrates that the temporal margin $\Delta \eta = 150$ is very effective at capturing  events where more oscillators are involved. Furthermore, the consistently low rate of \textit{Missed} events ($< 1\%$) confirms that cluster transitions in the adaptive Kuramoto model are predictable within an intermediate time scale.  Our approach relies exclusively on a single temporal snapshot of the system state. \\
What cannot be predicted is the new partition into two clusters after the switching event and the dynamics on a long-time scale.
One might be interested in the probability that a given  set of oscillator phases finds the same couplings (actually coupling conditions for their own evolution) after some possibly long time, as the evolution -locally in time- looks random. The question for finding repeatedly the same conditions is the question for long-time periodicity. Such long periods in the Kuramoto order parameter from short period oscillations have been observed in \cite{labavic2017long}. For larger networks we cannot answer this question, but for a system of $N=4$ oscillators, the six upper couplings and the six lower couplings synchronize into one or two clusters, as well as the phases, both for $N=4$ and $N=3$ the system has only one saddle point and the evolution of all couplings is strictly periodic, see Fig.~\ref{fig6}.

\vskip3pt
\noindent In summary, on intermediate time scales it is the momentary distance from $\pi$ of the phase position and the phase velocity at that instant of time, which allow a  prediction of which oscillator (group of oscillators) will escape next, knowing that there is an escape event. As soon as this distance is overcome, we can be sure to be in the attraction regime of the other cluster. It is the random initial conditions which determine the initial phase position and lead to an ensemble of time evolutions unless identical initial conditions are chosen via the same seed. Given a set of fixed initial conditions, the waiting time distribution between two switches of the order parameter $R_2$ looks random on longer time scales, unless it happens to become periodic. We address this irregularity to the complex attractor landscape that allows the escape from saddles and determines the entrance to the attraction regimes of new saddles. The entrance location determines the dwell time in the vicinity of the new saddles.

\begin{figure*}[]
	\centering
	\includegraphics[width=\linewidth]{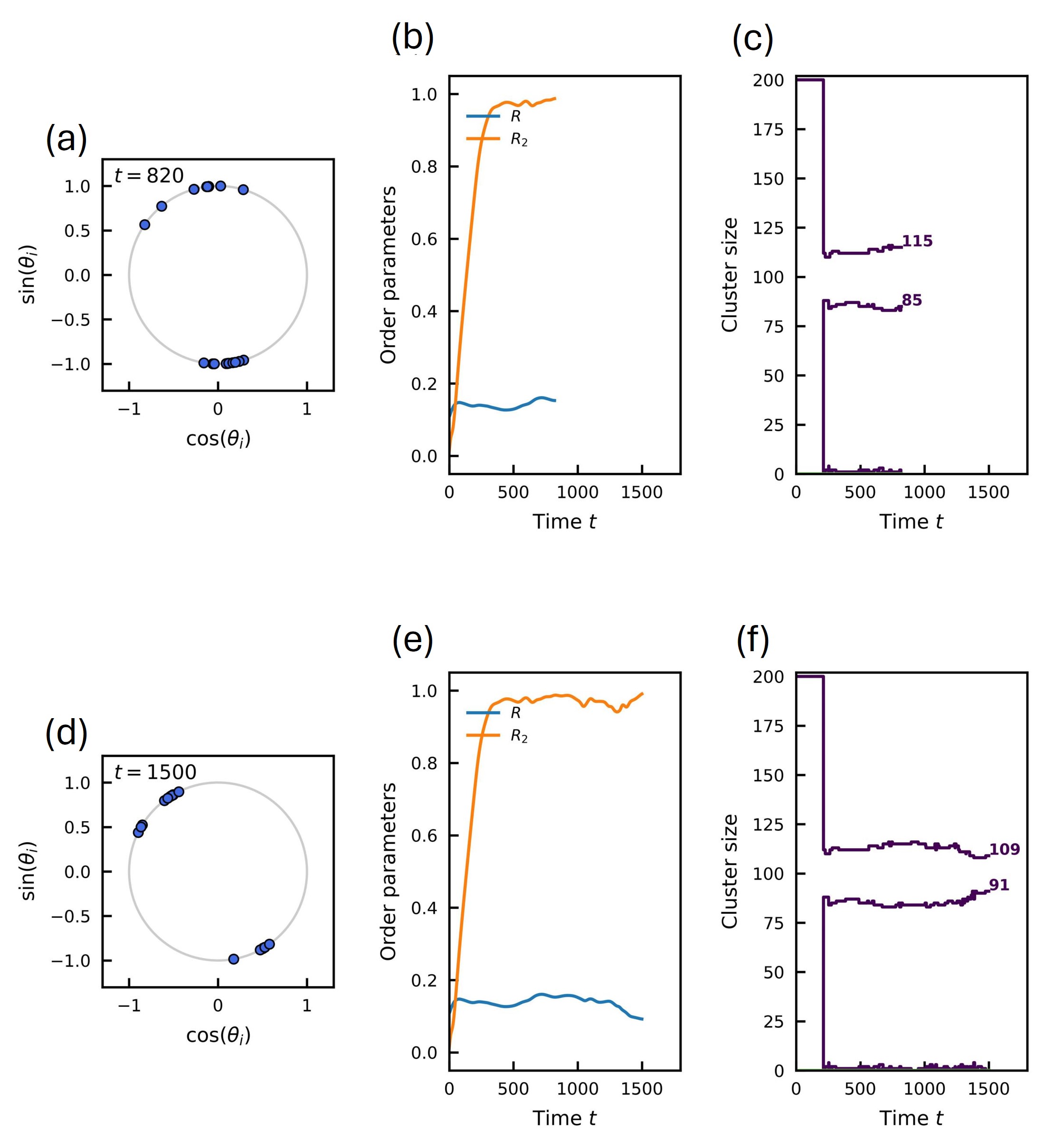}
	\caption{Snapshots of the system state before and after a switching event for the larger system size $N=200$. \textbf{(a--c)} State at $t=820$ (before the switch). (a) Instantaneous phases on the unit circle; note that the two clusters are less perfectly synchronized than in the $N=20$ case. (b) Time evolution of order parameters. (c) Cluster sizes showing a partition of $115$ and $85$ oscillators. \textbf{(d--f)} State at $t=1500$ (after the switch). (d) Instantaneous phases. (e) Evolution of order parameters. (f) Cluster sizes have shifted to a partition of $109$ and $91$ oscillators. Compared to $N=20$, the fluctuations in cluster size are more frequent, involving multiple oscillators switching affiliation.}
	\label{fig:switch_N200}
\end{figure*}

\begin{figure*}[htbp]
	\centering
	\includegraphics[width=0.8\textwidth]{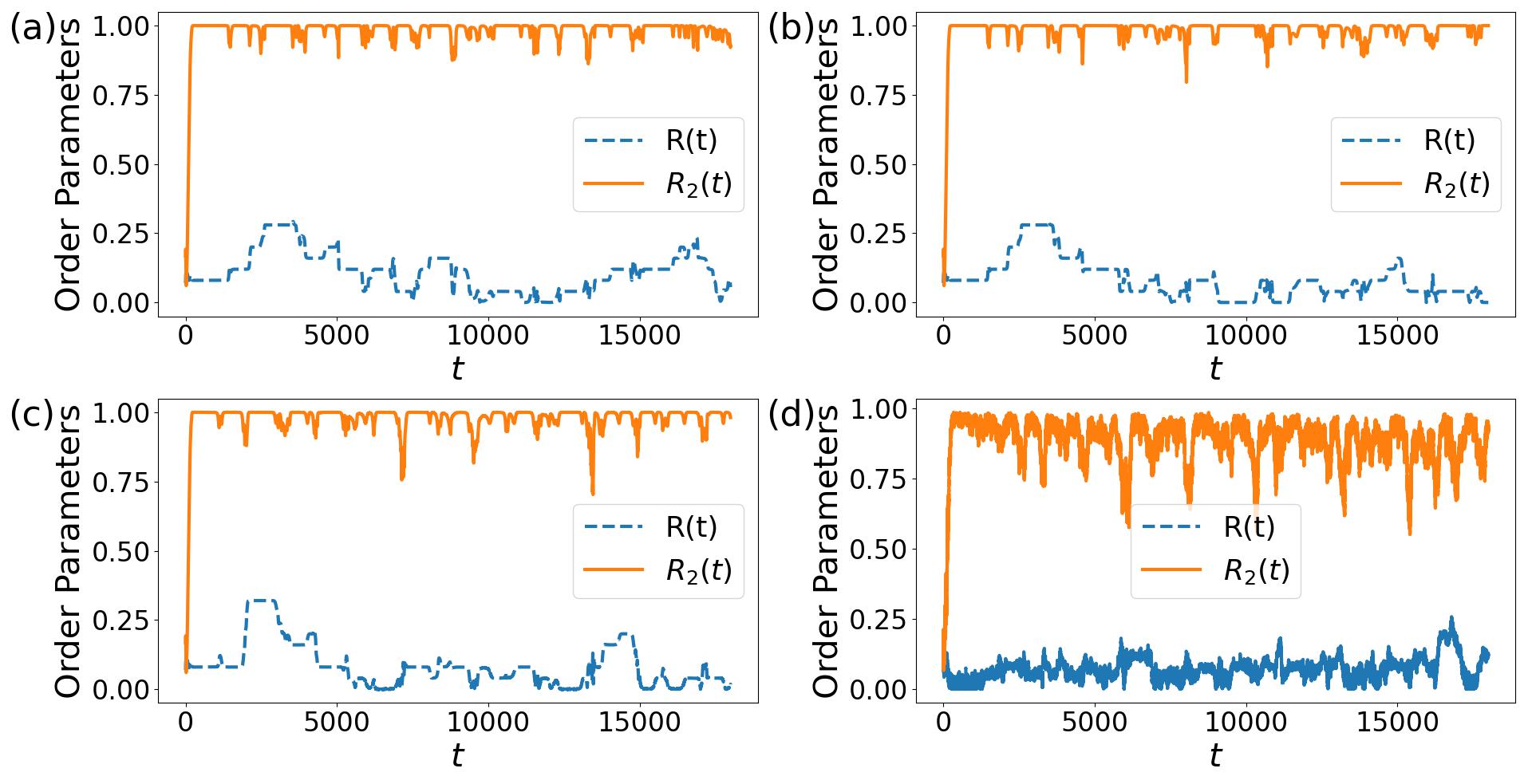}
	\caption{Time evolution of the order parameters $R$ and $R_2$ for four noise strengths: (a) $\sigma=0$, (b) $\sigma=0.00001$, (c) $\sigma=0.001$, (d) $\sigma=0.1$, leading to incoherent oscillations underlying $R$ and $R_2$ in (d). Parameters $N=50$, others as in Fig.~\ref{fig2}. We use the same random initial conditions to generate the four figures in panel (a)-(d). For further comments see the main text.
	}
	\label{fig11}
\end{figure*}

\subsection{Choice of initial conditions and natural frequencies} 
Before we discuss the stability properties as a function of the system size, some remarks are in order about the choice of initial conditions and the values of the natural frequencies.
If we keep all other parameters the same and set the intrinsic frequencies $\omega_i$ to a homogeneous nonzero value, e.g., $\omega_i = 3$, we observe that the phases $\theta_i$ oscillate over time 
with constant phase difference of $\pi$ between the temporarily forming two anti-phase synchronized clusters, otherwise the main features of metastability remain qualitatively the same as for $\omega_i=0$. This is expected as we can go to a co-rotating frame and transform the system to zero frequency.\\

If we choose uniformly distributed initial configurations for upper and lower couplings very close to 1, this choice results in some delay of the onset of metastable dynamics, where for the upper couplings it is  the fully synchronized state that remains stable until metastability sets in,  see Fig.~\ref{fig7}.

\begin{figure*}[htbp]
	\centering
	\includegraphics[width=0.6\textwidth]{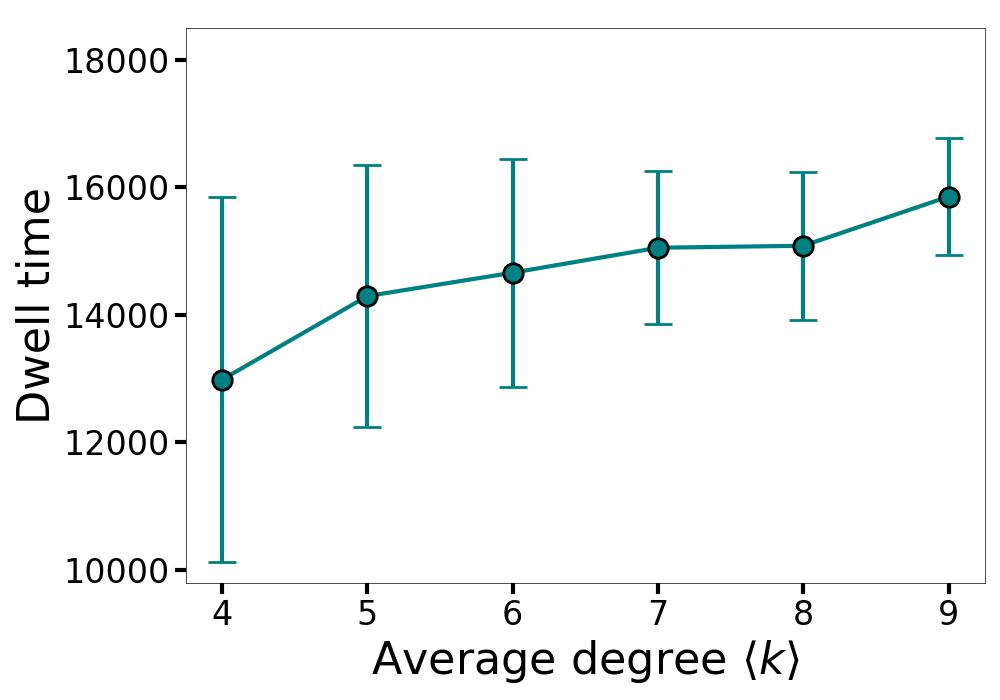}
	\caption{Increase of the dwell time in the vicinity of saddles with the network connectivity. The metastability is most pronounced for a fully connected network with average degree $\langle k\rangle=9$ and $N=10$.
	}
	\label{fig12}
\end{figure*}

\subsection{Metastability as a function of the system size}\label{subsechilde1}
In Fig.~\ref{fig8} we compare the evolution of the order parameters as a function of the system size for $N=10$ in Fig.~\ref{fig8} (a) and $N=200$ in Fig.~\ref{fig8} (b).

For $N=200$ the distinction between  temporarily stable configurations interrupted by short excursions to other clusters becomes less pronounced. The movie of \citep{Sayantan2025Metastability} for $N=200$ shows the evolution of phases, snapshots are given in Fig.\ \eqref{fig:switch_N200} before and after a spike in $R_2$. 	

We still see  remnants of an organization into two anti-phase synchronized clusters, where on short time scales always some oscillators escape and return to their cluster, more frequent are the events where the escape succeeds over a distance $\pi$ in the phases.
At intermediate times also configurations with several coexisting clusters are forming, leading to stronger spikes in $R_2$.\\
To understand more quantitatively the increasing instability with the system size, we summarize the stability analysis in the following . It is detailed in the Appendix. For the fully connected Kuramoto model with $\omega_i=0$ and $\sigma=0$ we start from a two-cluster anti-phase synchronized solution with clusters A and B of size $N_A$ and $N_B$. Next we perform a stability analysis around the two-cluster solution. The Jacobian takes a block form with four blocks $J_{\theta\theta}$, $J_{\theta k}$, $J_{k\theta}$, and $J_{kk}$. For the phase-coupling block $J_{\theta k}$ it is easy to show that the matrix vanishes. Therefore the determinant of the entire Jacobian factorizes into those of the diagonal entities $\det{J_{\theta\theta}}\det{J_{kk}}$, and the eigenvalues are the union of those of $J_{\theta\theta}$ and $J_{kk}$. The eigenvalues of $J_{kk}$ turn out to be all negative. Thus the problem is reduced to determine the eigenvalues of $J_{\theta\theta}$. At the two-cluster anti-phase equilibrium the couplings attain their equilibrium values $\dot{k}_{ij}=0$, which leads to vanishing row sums and one zero eigenvalue, corresponding to the global phase shift of the two-cluster solution. Moreover, it can be shown that $\chi J_{\theta \theta}\chi=-J_{\theta\theta}$ with a flip matrix $\chi$. Therefore the eigenvalues of $J_{\theta\theta}$ occur in symmetric pairs with $m-1$ positive values $\lambda_k=1-2k/N, k=1,2,...,m-1$  and $m-1$ negative values $\lambda_{-k}=-(1-2k/N), k=1,2,...,m-1$ for $N=2m$ plus $2$ zero eigenvalues;  for $N=2m+1$  we have $m$ positive and $m$ negative values and one zero eigenvalue, $\lambda_k, \lambda_{-k}$ as before with $k=1,2,...m$. This proves that with increasing system size $N$, we have an increasing number of unstable directions for the two anti-phase synchronized cluster solution.\\
\noindent What does this scaling mean in view of possible neural network applications? As stated in the introduction, individual Kuramoto oscillators are often used to model entire brain regions such as cortical areas, each treated as one oscillatory unit. For small system sizes such as $N=10$ or $N=20$ this could mean that one observes a pronounced metastability of synchronization patterns between brain areas. For a long dwell time they are synchronized in one pattern of two anti-phase synchronized clusters of areas, until suddenly  one or a few of these areas switch their cluster affiliation: after the switch, they synchronize with the  cluster  complementary to the one they were synchronized with before the switch. The dwell time in one synchronization pattern decreases with the number of coupled areas, the more areas, the shorter the dwell time. We leave it open which assignment to a single Kuramoto oscillator is most appropriate, ranging from a single neuron to global areas in the brain.

\subsection{Impact of disorder via noise or natural frequencies}
In Kuramoto models with two non-reciprocally interacting populations, disorder in the frequencies or noise can stabilize a chiral phase, which is chiral in the sense that instead of static phase alignment a chiral motion is observed such that in spite of zero natural frequencies, oscillators spontaneously rotate either clockwise or anti-clockwise \citep{fruchart2021non}. In our version of the Kuramoto model with one population and adaptive non-reciprocal couplings to all other oscillators, the role of disorder or noise that we observed is toward an increase of escapes from a two anti-phase synchronized cluster state. For the strongest noise amplitude in Fig.\ref{fig11} (d), the order parameter $R_2$ shows larger deviations from $1$, indicating that temporarily  several clusters have formed, but $R_2$ remains close to $1$ and returns to this value again and again, so that remnants of the anti-phase synchronized two clusters are still visible. Since the phase difference between the two clusters is roughly $\pi$, noise is in general not sufficient for an escape where many oscillators  overcome the distance and change their cluster affiliation at the same time, just a few succeed as the panels (c) in Fig.~\eqref{fig:switch_N200} show. In particular, noise does not improve synchronization at an intermediate noise level; instead, due to a large phase difference, it monotonically  weakens the perfect anti-phase cluster synchronization, allowing for longer escapes, and broadens a bit the in-phase synchronization within a cluster.

\begin{figure*}[htbp]
	\centering
	\includegraphics[width=1.0\textwidth]{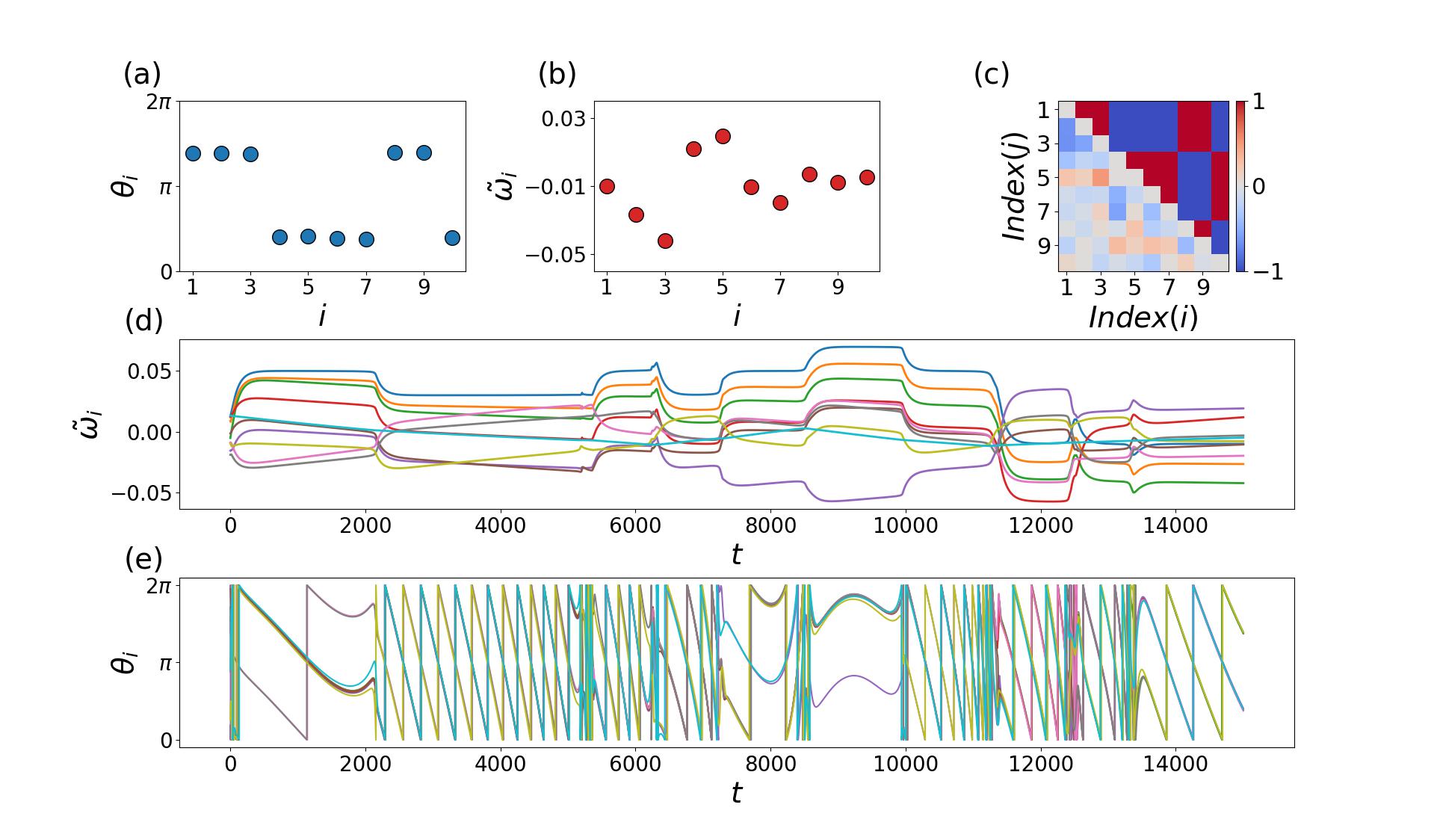}
	\caption{When frequencies become coupling dependent. Panel (a) shows a snapshot of the two anti-phase clusters of the oscillator phases. Panel (b) displays the instantaneous frequencies at the same snapshot time t=15000, revealing incoherent frequency dynamics. The evolution of the couplings is shown in panel (c), where the upper couplings rapidly reach their equilibrium values of ±1, while the lower couplings fail to do so. Panel (d) presents the time evolution of the frequencies over a longer time interval, and panel (e) shows the corresponding long-time evolution of the phases. Further details are discussed in the main text.}
	\label{fig13}
\end{figure*}

\subsection{Effects of a sparse network topology}
In the next version of the model, inspired by the Ref.\ \citep{kasatkin2018effect}, the dynamics of the oscillators and their adaptive couplings are modified to account for a connected but sparse random network topology. The coupling matrix $k_{ij}$ now evolves only along the links of a network represented by the adjacency matrix $A = [A_{ij}]$. The phase dynamics are therefore weighted by the network structure:
\begin{equation}
	\dot{\theta}_i = \omega_i + \frac{1}{\sum_j A_{ij}} \sum_{j=1}^{N} A_{ij} k_{ij} \sin(\theta_j - \theta_i), 
	\label{eq:theta_dynamics_sparse}
\end{equation}
where $\sum_j A_{ij}$ is the degree of node $i$.
The adaptive evolution of the coupling matrix is modified such that only  links, which are already initially existing, update according to the asymmetric  rules:
\begin{equation}
	\dot{k}_{ij} = 
	\begin{cases}
		-\varepsilon_1 \left[k_{ij} + \sin(\theta_i - \theta_j - \frac{\pi}{2})\right] A_{ij}, & i < j, \\
		-\varepsilon_2 \left[k_{ij} + \sin(\theta_i - \theta_j + \frac{\pi}{2})\right] A_{ij}, & i > j, \\
		0, & i = j.
	\end{cases}\label{eqnek}
\end{equation}

Using the dynamics~\eqref{eq:theta_dynamics_sparse}-\eqref{eqnek}, we simulated networks of size $N=10$ over multiple realizations for varying average degrees $\langle k \rangle$.  
The lifetime of the two-cluster anti-phase state was computed by monitoring the  order parameters $R$ and $R_2$, identifying periods where $R < R_{\mathrm{th}}$ while $R_2 > R_{\mathrm{th}}$.\\
Figure~\eqref{fig12} shows the mean and standard deviation of the two-cluster lifetimes over $100$ realizations as a function of the average degree $\langle k \rangle$.  As $\langle k \rangle$ increases and the network becomes more densely connected,  on average the dwell time increases and the metastability is more pronounced.  
The network topology introduces fluctuations in the dwell times across realizations, reflected in the error bars. Using our rough synchronization threshold of $0.8$, we do not see a pronounced rearrangement of clusters of the synchronized oscillators, differently from \citep{kasatkin2018effect}, in our case the two anti-phase cluster configuration is quite dominant.

\subsection{Kuramoto Model with non-reciprocal and adaptive frequencies}

Reciprocal synaptic plasticity alone can have the interesting effect that heterogeneous layered clusters with different frequencies emerge from homogeneous populations as the Fourier zero modes of the phase coupling function is included, leading to an additional constant term in the interaction part of the oscillators \citep{aoki2015self}. 
Here we implement a Kuramoto model,  inspired by the framework outlined in \citep{aoki2015self}, but with asymmetric plasticity. In this model, the frequency of each oscillator, denoted by $\tilde{\omega_i}$, depends on the coupling strengths, which evolve dynamically over time. The frequency of the $i$-th oscillator is given by the following equation:
\begin{equation}
	\tilde{\omega_i} = \omega_i + \frac{\Gamma_0}{N} \sum_{j=1}^{N} k_{ij},
\end{equation}
where $\Gamma_0 \in [0, 1]$ is a constant that influences the phase dynamics between oscillators. 
The phase evolution for each oscillator is governed by the equation:
\begin{equation}
	\frac{d\theta_i}{dt} = \omega_i + \frac{1}{N} \sum_{j=1}^N k_{ij} \left( \Gamma_0 - \sin(\theta_i - \theta_j + \alpha) \right),
	\label{eq:Aoki}
\end{equation}
where $\alpha$ is a constant that modulates the coupling strength between oscillators. The coupling matrix $k_{ij}$ now evolves according to the non-reciprocal plasticity rule as defined by Eq.~\eqref{eq:k_dynamics}, with $\beta_1 = - \frac{\pi}{2}$ and $\beta_2 = + \frac{\pi}{2}$. Initially, we set $\alpha = 0$, which simplifies the model to the noiseless version of the proposed model \eqref{eq:theta_dynamics} with $\Gamma_0 = 0$.\\
Even when the initial natural frequencies $\omega_i$ are set to zero, a nonzero $\Gamma_0$ ensures that the frequencies $\tilde{\omega_i}$ remain time-dependent. This leads to incoherent oscillations among frequencies. Furthermore, when $\Gamma_0$ is nonzero, the phases $\theta_i$ of the oscillators also evolve, and again we observe metastability in the system. This metastable behavior persists for values of $\Gamma_0$ in the range $[0, 0.2]$.\\
At small nonzero values of $\alpha \in [-0.04, 0.04]$, we observe the formation of two distinct clusters, with a phase difference of approximately $\pi$ between them. As $\Gamma_0$ increases to the range $(0.2, 0.6)$, the system still maintains two phase-locked clusters, but the phase difference between them is typically less than $\pi$. When $\Gamma_0$ exceeds $0.6$ and lies in the range $[0.6, 1]$, the oscillators become fully incoherent, and no phase locking is observed.\\
The time evolution of the order parameters and upper and lower couplings looks qualitatively the same as before. Whenever $R_2$
approaches $1$ (not displayed in Fig.\ \eqref{fig13}), the oscillators' phases naturally divide into two groups. Despite all oscillators starting with \( \omega_i = 0 \), they begin to oscillate due to the nonzero value of \( \Gamma_0 = 0.1 \).

We show a snapshot at \( t = 15000 \) of the phases and frequencies \( \tilde{\omega_i} \) in Fig.\ \eqref{fig13} (a) and (b), where each oscillator has a distinct \( \tilde{\omega_i} \) that oscillates over time. The 10 oscillators are divided into two groups separated by a phase difference of \( \pi \), although the frequencies are quite heterogeneous. 
In panel (c), we display a snapshot of the coupling matrix \( k_{ij} \). The lower triangle of the matrix does not yet converge to its equilibrium values of \( \pm 1 \) due to the slow adaptation rate \( \varepsilon_2 = 0.0001 \), the upper one does. 
One can easily read off which oscillators remain in the same cluster with upper couplings being attractive and which ones are in opposite clusters with upper couplings being repulsive. The long-time evolution of the phases $\theta_i$ (Fig.\ \eqref{fig13} (e)) indicates how the anti-phase synchronized clusters change together with the frequency  patterns (d) in time intervals between abrupt changes of the slope of $\tilde{\omega}_i$ and $\theta_i$. Note that $\tilde{\omega}_i$ play the role of effective natural frequencies. Obviously they can considerably differ from the actual frequencies $\dot{\theta}_i$, being larger in the time interval  between $t\in [2000,4000]$ than in  the one between $t\in [8000,10000]$, although the opposite order is seen for $\tilde{\omega}_i$.

\section{Conclusions and Outlook}
We have studied generalized Kuramoto models of classical oscillators which have applications in particular to model oscillatory dynamics in neuronal networks. Strengths or weights assigned to synapses are in general not symmetric, therefore it is natural to model their dynamics via non-reciprocal couplings, where we have chosen Hebbian and anti-Hebbian rules for plasticity. Moreover, the time scales on which adaptation takes place need not be the same in both directions. For a certain range of parameters and in this combination of asymmetric coupling's type and time-scale, we observe metastability of (in general) two anti-phase  clusters of synchronized oscillators where a few oscillators (``switcher oscillators") change their cluster affiliation. This is induced by a dynamical switch of the coupling type (attractive or repulsive) of connections $ij$ to nodes $j$ from  switcher oscillators $i$, which happen on different time scales, depending on whether $i<j$ or $i>j$. We see temporary remnants of dynamical realizations of Harary's theorem, that is, the existence of two anti-phase synchronized clusters on an all-to-all topology, as long as the dynamics on the second time scale, parameterized by $1/\epsilon_2$, can be neglected.\\
Regarding the system size, individual oscillators may represent mesoscopic neural populations such as cortical columns \citep{roberts2019metastable}. Therefore a small system size of $N=10$ or $N=20$ with pronounced metastability would include many neurons; here the increasing instability for large $N$, derived from our scaling analysis, would not apply to the individual neurons, but to a large number of brain areas.   The `switcher' mechanism would mean that some areas are flexible with respect to the choice of the  cluster with which  to synchronize. In view of experiments, first steps may amount to identify interactions which are Hebbian-like in one direction and anti-Hebbian-like in the opposite direction together with the time scales they are acting on.\\
In our current model with non-reciprocal phase shifts of $\pm \pi/2$, the system  favors two anti-phase clusters. Different options exist to obtain states with more than two clusters or different collective behaviors (e.g., a splay state \citep{berner2021generalized,singha2016spatial}, or generalized chimeras \citep{parastesh2021chimeras, khaleghi2019chimera}). One would typically need to modify the interaction function to include higher-order interactions \citep{kundu2022higher} or alter the phase lag parameters \citep{kasatkin2017self} away from strict orthogonality, as shown in previous studies on clustering in coupled oscillator networks \citep{okuda1993mutual, hansel1993phase}. An open question is as to whether switchers between more than two clusters could be created due to the same kind of non-reciprocal adaptive couplings. Finally,  in order to see new collective phenomena in neural networks, one may  search for dynamical phases as result of non-reciprocal interactions in models of limit-cycle oscillators such as Stuart-Landau or Fitz-Hugh Nagumo oscillators, including amplitudes apart from oscillator phases as considered in this work.

\begin{figure*}[htbp]
	\centering
	\includegraphics[width=0.8\textwidth]{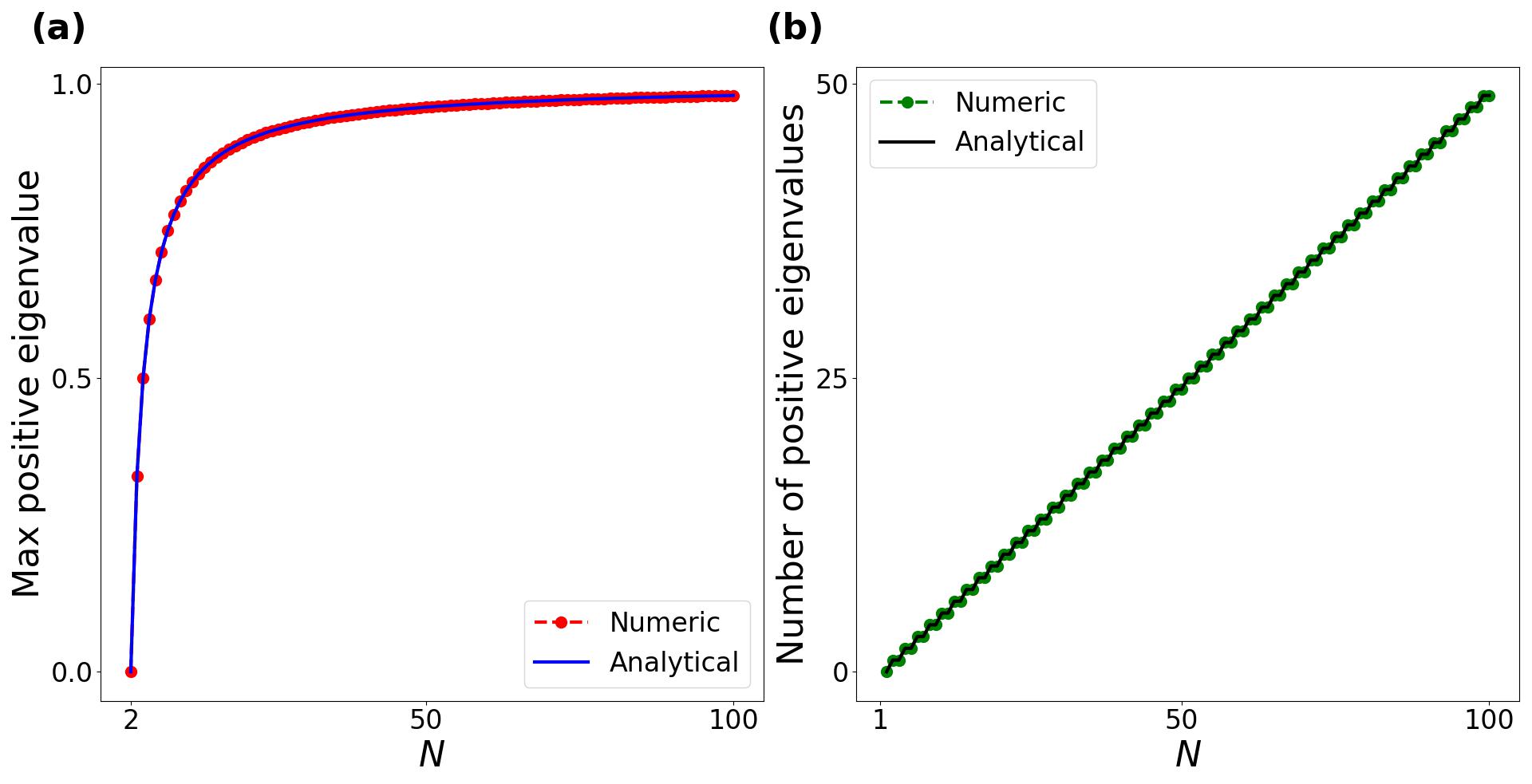}
	\caption{
		Stability analysis of the two-cluster anti-phase state using the phase-phase block $J_{\theta\theta}$ (given in Eq.\ \eqref{eq:J_tt_entries}) of the Jacobian. 
		\textbf{(a)} Maximum positive eigenvalue as a function of $N$.
		\textbf{(b)} Maximum number of positive eigenvalues as a function of the number of oscillators $N$. 
		The results confirm that the numerical maxima agree with the analytical predictions. 
	}
	\label{fig14}
\end{figure*}

\appendix
\section{Appendix: Stability analysis of two anti-phase clusters of arbitrary sizes}
\label{sec:appendix_stability_analysis}
Consider the original fully connected Kuramoto model \eqref{eq:theta_dynamics} in the absence of noise ($\sigma_{\mathrm{noise}} = 0$) with $\omega_i  = 0$ $\forall$ $i=1,2,3,\cdots,N$.  
Assume the oscillators form two anti-phase clusters, named Cluster A and Cluster B, of sizes $N_A$ and $N_B = N - N_A$, with phases
\begin{equation}
	\theta_i = 
	\begin{cases}
		\theta_A, & i \in \text{Cluster A},\\
		\theta_B = \theta_A + \pi, & i \in \text{Cluster B}.
	\end{cases}
\end{equation}
The state vector is
\begin{equation}
	\mathbf{x} = (\theta_1, \dots, \theta_N, k_{ij} \text{ for } i \neq j) \in \mathbb{R}^{N + N(N-1)},
\end{equation}
where we exclude the fixed diagonal elements $k_{ii} = 0$.
Linearizing the dynamics around the two-cluster solution, the Jacobian takes the four-block form:
\begin{equation}
	\mathbf{J} =
	\begin{pmatrix}
		\mathbf{J}_{\theta\theta} & \mathbf{J}_{\theta k} \\
		\mathbf{J}_{k \theta} & \mathbf{J}_{k k}
	\end{pmatrix}.
\end{equation}
with  $\mathbf{J}_{\theta\theta}$: $N \times N$, phase--phase interactions;
$ \mathbf{J}_{\theta k}$: $N \times N(N-1)$, effect of couplings on phases;
$ \mathbf{J}_{k \theta}$: $N(N-1) \times N$, effect of phases on couplings;
$ \mathbf{J}_{kk}$: $N(N-1) \times N(N-1)$, coupling–coupling interactions.\\
Let us denote the blocks of the Jacobian as follows:\\
{\bf Phase–Phase block: $\mathbf{J}_{\theta\theta}$}
The entries are
\begin{equation}
	(\mathbf{J}_{\theta\theta})_{ij} = \frac{\partial \dot{\theta}_i}{\partial \theta_j} \\
	= \frac{\partial}{\partial \theta_j} \left[ \omega_i + \frac{1}{N} \sum_{m=1}^{N} k_{im} \sin(\theta_m - \theta_i) \right].
\end{equation}
Evaluating the derivative gives
\begin{equation}
	(\mathbf{J}_{\theta\theta})_{ij} =
	\begin{cases}
		-\frac{1}{N} \sum\limits_{m \neq i} k_{im} \cos(\theta_m - \theta_i), & i=j,\\[1mm]
		\frac{1}{N} k_{ij} \cos(\theta_j - \theta_i), & i \neq j.
	\end{cases}
\end{equation}
{\bf Phase–Coupling Block: $\mathbf{J}_{\theta k}$}
This block captures how changes in couplings affect the phase dynamics:
\begin{equation}
	(\mathbf{J}_{\theta k})_{i,(j\neq l)} = \frac{\partial \dot{\theta}_i}{\partial k_{jl}} =
	\begin{cases}
		\frac{1}{N} \sin(\theta_l - \theta_i), & j=i,\\
		0, & j \neq i.
	\end{cases}
\end{equation}
{\bf Coupling–Phase Block: $\mathbf{J}_{k\theta}$}
This block gives how the phases affect the evolution of couplings:

\begin{equation}
	\begin{aligned}
		(\mathbf{J}_{k \theta})_{(i\neq j), m} 
		&= \frac{\partial \dot{k}_{ij}}{\partial \theta_m} \\
		&= 
		\begin{cases}
			- \varepsilon_1 \cos(\theta_i - \theta_j - \pi/2), & i<j, \, m=i,\\
			\varepsilon_1 \cos(\theta_i - \theta_j - \pi/2), & i<j, \, m=j,\\
			- \varepsilon_2 \cos(\theta_i - \theta_j + \pi/2), & i>j, \, m=i,\\
			\varepsilon_2 \cos(\theta_i - \theta_j + \pi/2), & i>j, \, m=j,\\
			0, & \text{otherwise.}
		\end{cases}
	\end{aligned}
\end{equation}

{\bf Coupling–Coupling Block: $\mathbf{J}_{kk}$}
Finally, the effect of couplings on themselves is diagonal:

\begin{equation}
	\begin{aligned}
		(\mathbf{J}_{kk})_{(i\neq j),(l\neq m)} 
		&= \frac{\partial \dot{k}_{ij}}{\partial k_{lm}} \\
		&= 
		\begin{cases}
			- \varepsilon_1, & i<j \text{ and } (i,j) = (l,m),\\
			- \varepsilon_2, & i>j \text{ and } (i,j) = (l,m),\\
			0, & \text{otherwise.}
		\end{cases}
	\end{aligned}
\end{equation}

{\bf Phase–Coupling Block is Zero}
For the two-cluster configuration, each oscillator in cluster A has phase \(\theta_A\), and each in cluster B has \(\theta_B = \theta_A + \pi\).  
Thus, for any entry of the phase–coupling block
\[
(\mathbf{J}_{\theta k})_{i,(j \neq l)} = 
\frac{\partial \dot{\theta}_i}{\partial k_{jl}} =
\begin{cases}
	\frac{1}{N} \sin(\theta_l - \theta_i), & j = i,\\
	0, & j \neq i,
\end{cases}
\]
we have two possibilities:
\begin{itemize}
	\item If \(\theta_l\) and \(\theta_i\) belong to the same cluster, then \(\theta_l - \theta_i = 0 \implies \sin(\theta_l - \theta_i) = 0\).
	\item If \(\theta_l\) and \(\theta_i\) belong to different clusters, then \(\theta_l - \theta_i = \pm \pi \implies \sin(\theta_l - \theta_i) = 0\) as well.
\end{itemize}
\noindent
Hence, for all \(i, j, l\),
\[
(\mathbf{J}_{\theta k})_{i,(j \neq l)} = 0 \quad \Rightarrow \quad
\mathbf{J}_{\theta k} = \mathbf{0}_{N \times N(N-1)}.
\]

\noindent
{\bf Eigenvalues via block-triangular structure}
For the two-cluster state, we have shown that the phase–coupling block vanishes:
\[
\mathbf{J}_{\theta k} = \mathbf{0}_{N \times N(N-1)}.
\]
Hence, the Jacobian becomes block lower-triangular:
\[
\mathbf{J} =
\begin{pmatrix}
	\mathbf{J}_{\theta\theta} & \mathbf{0} \\
	\mathbf{J}_{k \theta} & \mathbf{J}_{kk}
\end{pmatrix}.
\]
Recall that the determinant of a block-triangular matrix is the product of the determinants of the diagonal blocks:
\[
\det
\begin{pmatrix}
	\mathbf{A} & \mathbf{0} \\
	\mathbf{C} & \mathbf{B}
\end{pmatrix}
= \det(\mathbf{A}) \cdot \det(\mathbf{B}),
\]
and applying this to \(\mathbf{J}\) gives the characteristic polynomial:
\begin{equation}
	\det(\mathbf{J} - \lambda \mathbf{I}) =  \, 
	\det(\mathbf{J}_{\theta\theta} - \lambda \mathbf{I}_N) 
	\cdot \det(\mathbf{J}_{kk} - \lambda \mathbf{I}_{N(N-1)}).
\end{equation}
\noindent
By definition, the eigenvalues are the roots of \(\det(\mathbf{J} - \lambda \mathbf{I}) = 0\).  
From the above factorization, the eigenvalues of \(\mathbf{J}\) are exactly the eigenvalues of the two diagonal blocks:
\[
\mathrm{eig}(\mathbf{J}) = \mathrm{eig}(\mathbf{J}_{\theta\theta}) \cup \mathrm{eig}(\mathbf{J}_{kk}).
\]
This proves that, due to the vanishing phase–coupling block, the spectrum of the full Jacobian splits into the spectra of \(\mathbf{J}_{\theta\theta}\) and \(\mathbf{J}_{kk}\).\\
\noindent {\bf Reduction to the Phase–Phase Block}
From the previous discussion, we have
\[
\mathbf{J}_{kk} = \text{diag}(-\varepsilon_1, \dots, -\varepsilon_1, -\varepsilon_2, \dots, -\varepsilon_2),
\]
where each diagonal entry is strictly negative. Therefore, all eigenvalues of \(\mathbf{J}_{kk}\) satisfy
$\lambda(\mathbf{J}_{kk}) < 0.$ 
Since the full Jacobian \(\mathbf{J}\) is block lower-triangular,
$\mathrm{eig}(\mathbf{J}) = \mathrm{eig}(\mathbf{J}_{\theta\theta}) \cup \mathrm{eig}(\mathbf{J}_{kk})$,
any positive eigenvalue of \(\mathbf{J}\) can only arise from \(\mathbf{J}_{\theta\theta}\).  
\noindent\textbf{Conclusion:} The stability of the two-cluster state is entirely determined by the \(N \times N\) phase–phase block \(\mathbf{J}_{\theta\theta}\).  
The \(N(N-1) \times N(N-1)\) coupling–coupling block \(\mathbf{J}_{kk}\) cannot generate any instability because it is diagonal with strictly negative entries.  
Hence, the analysis reduces to checking the eigenvalues of \(\mathbf{J}_{\theta\theta}\) only, greatly simplifying the stability problem.\\

\noindent{\bf Properties of the Phase–Phase Block \(\mathbf{J}_{\theta\theta}\)}
The entries of \(\mathbf{J}_{\theta\theta}\) are real:
\[
(\mathbf{J}_{\theta\theta})_{ij} =
\begin{cases}
	-\frac{1}{N} \sum\limits_{m \neq i} k_{im} \cos(\theta_m - \theta_i), & i=j,\\[1mm]
	\frac{1}{N} k_{ij} \cos(\theta_j - \theta_i), & i \neq j.
\end{cases}
\]
{\bf Row sum and zero eigenvalue:}  Observe that for each row \(i\), the sum of entries is

\begin{equation*}
	\begin{aligned}
		\sum_{j=1}^{N} (\mathbf{J}_{\theta\theta})_{ij} 
		&= -\frac{1}{N} \sum_{m \neq i} k_{im} \cos(\theta_m - \theta_i) \\
		&\quad + \sum_{j \neq i} \frac{1}{N} k_{ij} \cos(\theta_j - \theta_i) \\
		&= 0.
	\end{aligned}
\end{equation*}

Therefore, \(\mathbf{J}_{\theta\theta}\) always has one eigenvalue exactly equal to zero:
$\lambda_0 = 0$.
This zero eigenvalue corresponds to the rotational invariance of the Kuramoto model. Shifting all phases by a constant amount
\(\theta_i \mapsto \theta_i + \phi_0\) does not change the dynamics, so perturbations along this uniform phase shift direction neither grow nor decay. Hence, the zero eigenvalue represents a neutral mode associated with the global phase of the two-cluster solution.

\noindent{\bf Symmetry of eigenvalues.}
At the two-cluster anti-phase equilibrium, the couplings attain their asymptotic values:
\begin{equation}
	k_{ij}^{\mathrm{eq}} =
	\begin{cases}
		- \sin(\theta_i - \theta_j - \pi/2), & i < j,\\
		- \sin(\theta_i - \theta_j + \pi/2), & i > j,
	\end{cases}
	\label{eq:k_eq}
\end{equation}
where $k_{ii}^{\mathrm{eq}} = 0$. 
Using these equilibrium couplings, the entries of the phase–phase Jacobian become
\begin{equation}
	(\mathbf{J}_{\theta\theta})_{ij} =
	\begin{cases}
		-\frac{1}{N} \sum\limits_{m \neq i} k_{im}^{\mathrm{eq}} \cos(\theta_m - \theta_i), & i=j,\\[1mm]
		\frac{1}{N} k_{ij}^{\mathrm{eq}} \cos(\theta_j - \theta_i), & i \neq j,
	\end{cases}
	\label{eq:J_tt_entries}
\end{equation}
and one can verify that the row sums vanish:
$\sum_{j=1}^{N} (\mathbf{J}_{\theta\theta})_{ij} = 0, \quad \forall i$.
Hence, the Jacobian $\mathbf{J}_{\theta\theta}$ always has at least one zero eigenvalue, corresponding to the uniform phase shift mode:
$\lambda_0 = 0$.
All entries of $\mathbf{J}_{\theta\theta}$ are real. Let us define the exchange (or flip) matrix $\chi \in \mathbb{R}^{N\times N}$ as
\begin{equation}
	\chi =
	\begin{pmatrix}
		0 & 0 & \cdots & 0 & 1\\
		0 & 0 & \cdots & 1 & 0\\
		\vdots & \vdots & \reflectbox{$\ddots$} & \vdots & \vdots\\
		0 & 1 & \cdots & 0 & 0\\
		1 & 0 & \cdots & 0 & 0
	\end{pmatrix}.
\end{equation}
This matrix is self-inverse, $\chi^2 = I_N$, and it satisfies the relation
$\chi \, \mathbf{J}_{\theta\theta} \, \chi = - \mathbf{J}_{\theta\theta}$.
As a consequence, since $\chi \, \mathbf{J}_{\theta\theta} \, \chi = - \mathbf{J}_{\theta\theta}$ is a similarity transformation, the eigenvalues of $\mathbf{J}_{\theta\theta}$ occur in symmetric pairs: if $\lambda$ is an eigenvalue, then $-\lambda$ is also an eigenvalue. This property is useful for analyzing the stability of the two-cluster solution.

\noindent{\bf Eigenvalue spectrum of $\mathbf{J}_{\theta\theta}$:}  
The phase–phase block $\mathbf{J}_{\theta\theta}$ always has at least one zero eigenvalue corresponding to the uniform phase shift mode. Combining this with the eigenvalue symmetry under the exchange matrix $\chi$, we can characterize the spectrum as follows:
\begin{itemize}
	\item If $N = 2m+1$ is odd, $\mathbf{J}_{\theta\theta}$ has one zero eigenvalue, $m$ positive eigenvalues, and $m$ negative eigenvalues. Moreover, the positive and negative eigenvalues have equal absolute values.
	\item If $N = 2m$ is even, $\mathbf{J}_{\theta\theta}$ has two zero eigenvalues, $m-1$ positive eigenvalues, and $m-1$ negative eigenvalues, again with matching absolute values.
\end{itemize}
Hence, as $N$ increases, the number of positive eigenvalues grows, indicating an increasing number of unstable directions for the two-cluster solution.

\noindent{\bf Eigenvalues of $\mathbf{J}_{\theta\theta}$:}  
By explicitly solving $\mathbf{J}_{\theta\theta}$ for the two-cluster configuration, one finds that the positive eigenvalues take the form
\begin{equation}
	\lambda_k = 1 - \frac{2k}{N}, \qquad k = 1, 2, \dots, n_+,
\end{equation}
where $n_+$ is the number of positive eigenvalues. The corresponding negative eigenvalues are
\begin{equation}
	\lambda_{-k} = -\lambda_k = -\left(1 - \frac{2k}{N}\right),  k = 1, 2, \dots, n_-,
\end{equation}
with $n_+ = n_-$, reflecting the eigenvalue symmetry of $\mathbf{J}_{\theta\theta}$.
Thus, for general $N$, the eigenvalues of $\mathbf{J}_{\theta\theta}$ consist of symmetric positive and negative pairs, along with one or two zero eigenvalues depending on whether $N$ is odd or even. As $N$ increases, the number of positive eigenvalues increases, indicating that the two-cluster configuration becomes less stable in larger networks.

\noindent
{\bf Numerical Validation:} 
To investigate the local stability of the two anti-phase clusters, we analyze the phase-phase block of the Jacobian, $J_{\theta\theta}$ (given in Eq.\ \eqref{eq:J_tt_entries}), corresponding to a given bipartition of $N$ oscillators. For each bipartition, we construct the equilibrium phases $\theta_i = 0$ for one cluster and $\theta_i = \pi$ for the other, and compute the real parts of the eigenvalues of $J_{\theta\theta}$. Due to the large number of possible bipartitions, a full enumeration is computationally infeasible. Indeed, the total number of distinct bipartitions for a system of $N$ oscillators is 
\[
\frac{1}{2} \sum_{k=1}^{N-1} \binom{N}{k} = 2^{N-1}-1,
\]
which for $N=20$ amounts to $524{,}287$ possible configurations. Therefore, we adopt a random sampling strategy, selecting 500 random bipartitions for each $N$. This approach allows us to efficiently approximate the maximal number of positive eigenvalues and the largest positive eigenvalue while keeping the computational cost reasonable. The results of this procedure are summarized in Fig.~\eqref{fig14}, which shows the maximal positive eigenvalue and the number of positive eigenvalues as functions of $N$. Clearly, as $N$ increases, both the number of positive eigenvalues and the maximum positive eigenvalue grow, with the latter approaching the theoretical limit of $1$.

\vspace*{1cm}

\section*{Conflict of Interest Statement}
The authors declare that the research was conducted in the absence of any commercial or financial relationships that could be construed as a potential conflict of interest.

\section*{Author Contributions}

The project was conceived by H.M.-O. The simulations and analytical calculations were performed by S.N.C. Both authors contributed to a first draft of the manuscript. H.M.-O. is responsible for the final version. Both authors agree about the final version.

\section*{Funding}
Financial support by the Deutsche Forschungsgemeinschaft (DFG) through
Grant No. ME-1332/30-1 is gratefully acknowledged. Furthermore, the simulations were performed on a computer cluster funded
through the project INST 676/7-1 FUGG.

\section*{Acknowledgments}
We would like to thank Dr. Tuan Minh Pham (University of Amsterdam) for stimulating discussions. 


\section*{Data Availability Statement}
The original contributions presented in the study are included in the article. The code used to generate the key findings and animations are publicly available in the GitHub repository \citep{Sayantan2025Metastability}.

\bibliographystyle{unsrt} 
\bibliography{library}

\end{document}